\documentclass[11pt]{article}

\usepackage[authoryear]{natbib}

\usepackage{amsmath}
\usepackage{graphicx}
\usepackage[colorinlistoftodos]{todonotes}
\usepackage[colorlinks=true, allcolors=black]{hyperref}
\usepackage{amsmath, amssymb,amsthm,bm,bbm,fullpage}
\usepackage{mathrsfs}
\usepackage{cite}
\usepackage{colortbl}
\usepackage[all]{xy}
\usepackage{epsfig}
\usepackage{subfigure}
\usepackage{graphics}
\usepackage{multirow}
\usepackage[affil-it]{authblk}
\usepackage{lineno}
\usepackage{rotating}

\usepackage{setspace}
\usepackage[font=footnotesize]{caption}

\newenvironment{sciabstract}{%
\begin{quote} \bf}
{\end{quote}}

\title{\textbf{Polarization under rising inequality and economic decline}} 

\author[1 *]{Alexander J. Stewart}
\author[2 *]{Nolan McCarty}
\author[3 *]{Joanna J. Bryson}

\affil[1]{Department of Biology, University of Houston, Houston, TX, USA}
\affil[2]{Woodrow Wilson School, Princeton University, Princeton, NJ, USA}
\affil[3]{Centre for Digital Governance, Hertie School, Berlin, Germany}
\affil[*]{astewar6@central.uh.edu, nmccarty@princeton.edu, jjb@alum.mit.edu}

\begin{document}
\date{}
\maketitle

\begin{sciabstract}
 Social and political polarization is a significant source of conflict and poor governance in many societies. Thus,  understanding its causes has become a priority of scholars across many disciplines. Here we demonstrate that shifts in socialization strategies analogous to political polarization and identity politics can arise as a locally-beneficial response to both rising wealth inequality and economic decline. Adopting a perspective of cultural evolution, we develop a framework to study the emergence of polarization under shifting economic environments. 
In many contexts, interacting with diverse out-groups confers benefits from innovation and exploration greater than those that arise from interacting exclusively with a homogeneous in-group.
However, when the economic environment favors risk-aversion, a strategy of seeking low-risk interactions can be important to maintaining individual solvency. 
To capture this dynamic, we assume that in-group interactions have a lower expected outcome, but a more certain one. Thus in-group interactions are less risky than out-group interactions.
Our model shows that under conditions of economic decline or increasing wealth inequality, some members of the population benefit from adopting a risk-averse, in-group favoring strategy.
Moreover, we show that such in-group polarization can spread rapidly to the whole population and persist even when the conditions that produced it have reversed.
Finally we offer empirical support for the role of income inequality as a driver of affective polarization in the United States, mirroring findings on a panel of developed democracies.
Our work provides a framework for studying how disparate forces interplay, via cultural evolution, to shape patterns of identity, and unifies what are often seen as conflicting explanations for political polarization: identity threat versus economic anxiety. 
\end{sciabstract}

\doublespace

\section*{Introduction}

The emergence of `populist' movements in countries as varied as  the USA,  UK, Brazil, Hungary, Poland, India and the Philippines, has left scholars, journalists, and other observers scrambling to understand the source of their support. Often discourse has been reduced to a horse race, pitting arguments focusing on social identity, such as racial, ethnic, and nationalistic hostilities, against those concerning the economic anxieties of populist movement supporters and parties. 

Adherents of both claims can find support for their arguments. Proponents of racial anxiety can offer cross-sectional and experimental evidence showing a connection between support for populist positions in the US and UK, and racial anxiety \citep{Schaffner:2016,Luttig:2017,Sides:2017,Inglehart:2016,Tesler16}, while advocates of economic anxieties can point to negative longer-term trends in the economic and social well-being of middle class voters \citep{Arnorsson:2016,Kolko16}, and its known correlation with polarized sentiments \citep{Mitrea20}.

We suggest that these arguments should be viewed as complementary rather than as competing.  Declines in economic well-being and social status that may result from inequality and economic decline may also induce changes in social behavior that trigger intra-group conflict along available cleavages. Journalistic observers have noted the complementarities between economic and racial anxiety before \citep{Atkins17,Levitz17,Robinson17,Judis:2016}, while a significant body of research has described the empirical relationships between legislative and affective polarization and inequality \citep{McCarty16,Gidron}. However, formal models that describe a mechanism by which polarization can arise in response to economic hardship have not been developed. 


Here we address this gap by developing a formal model for the dynamics of in- and out-group interactions under a changing economic environment. Adopting the framework of cultural evolution, we assume an individual's economic success is determined both by her interactions with others and by the underlying state of the economy. Further we assume that the behavior of successful individuals is likely to be emulated and expressed by others.
In this context, we examine outcomes for a socially-acquired behavioral strategy that encodes each individual's choice of whether to interact in a large population with those who are `like' (in-group interactions) or `unlike' (out-group interactions) the self in a variety of situations of economic change. 

In our model, reliance on in-group interactions is assumed to be less risky but entails lower rewards for success, compared to a strategy of reliance on out-group interactions.

A range of empirical results show the general benefits of diversity on successful decision making \citep{ruef2002strong,woolley2010evidence,Martinez11} --- intuitively, interactions with more diverse out-group members may be assumed to pool greater knowledge applicable to a wider variety of situations. Such interactions, when successful, generate better solutions and greater benefits. 
However, we also assume that there is increased risk of failure for out-group interactions, due to a weaker capacity to coordinate among individuals, compared to more familiar in-group interactions \citep{Carruthers96}. 


We show that under a broad range of conditions, the trade-off between risk reduction and benefit maximization decreases out-group interactions when a population is faced with economic decline. We show that such group polarization can be contagious, and a sub-population facing economic hardship in an otherwise strong economy can tip the whole population into a state of sub-optimal polarization. Similarly, we show that a population that becomes polarized can remain trapped in that state, even after a reversal of the conditions which generated the risk aversion and polarization in the first place. 

Finally we examine the empirical relationship between inequality in the USA, and affective party polarization ---  a survey measure of the mutual dislike of out-partisans which has been posited to be related to the social group cleavages of the party system \citep{Iyengar:2012,Iyengar:2015,Mason:2015,Mason:2018}. 
Using data from the last three presidential election cycles drawn from the American National Election Study (ANES) and the Cooperative Congressional Election Study (CCES), we show that inequality and affective polarization are correlated across US states, consistent with similar findings that inequality and affective polarization are correlated in a panel of developed democracies. This evidence provides further support for our theoretical framework. Taken together our work offers both a theoretical account and empirical support for the emergence of polarization as a response to economic hardship.  Our framework also suggests an explanation for the apparent difficulty in reversing polarization once it becomes established \citep{McCarty16}. 
\\

\begin{figure}[th] \centering \includegraphics[scale=0.4]{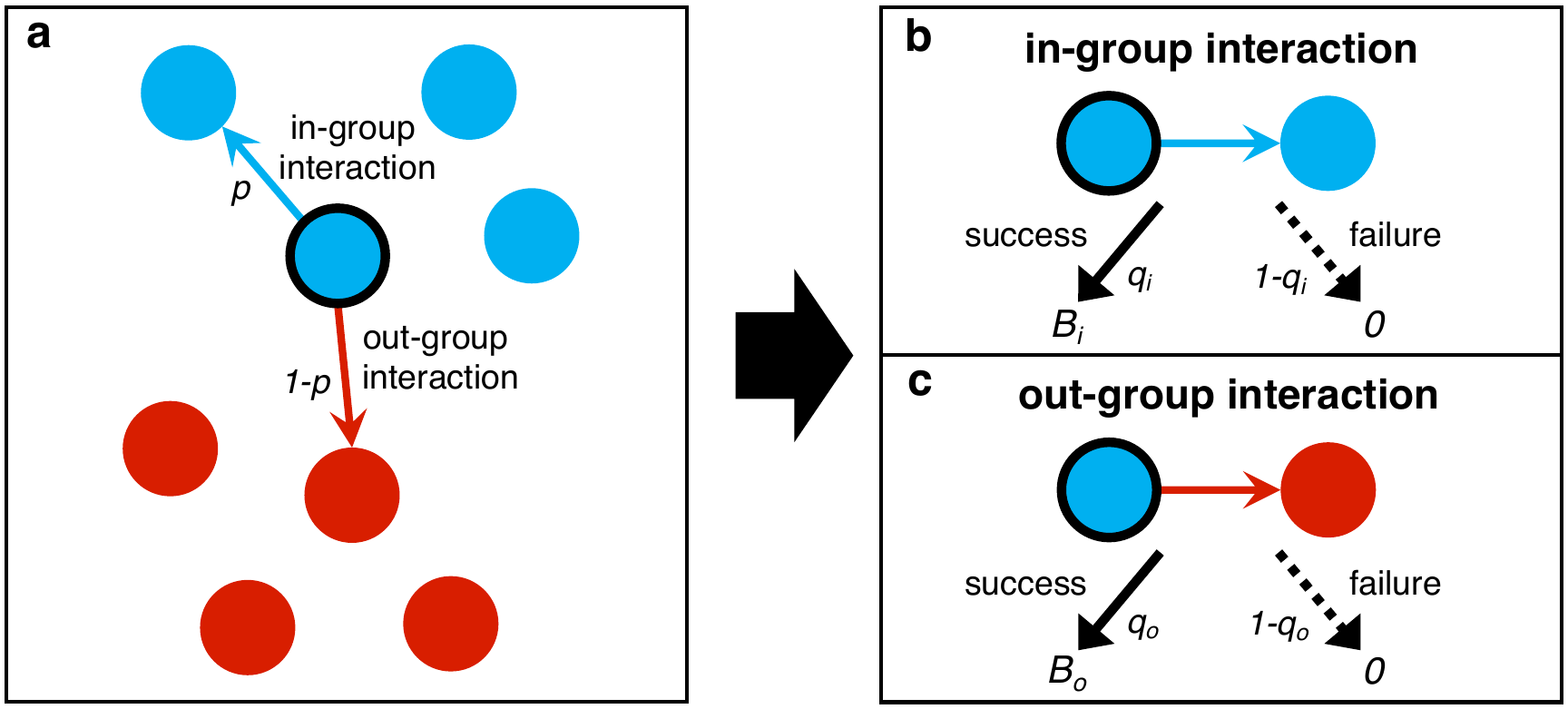}
\caption{A simple model of diversity in social interactions. a) From the perspective of a focal individual (black outline) a population is divided into players who are `like' self (in-group, here blue) and `unlike' self (out-group, here red).  Each player has a strategy such that in any given interaction, they choose either a member of the in-group with probability $p$ (blue arrow) or the out-group with probability $1-p$ (red arrow). b) If the focal player participates in an in-group interaction, it is successful with probability $q_i$ generating a benefit $B_i>0$. Otherwise the interaction fails with probability $1-q_i$ and generates no benefit. c) Similarly, if the focal player participates in an out-group interaction it succeeds with probability $q_o$ generating a benefit $B_o>0$. Conversely, the interaction fails with probability $1-q_o$ and  generates no benefit. Finally we assume that out-group interactions are associated with higher rewards $B_o>B_i$ and higher risks $q_o<q_i$. Note that in our models, groups are arbitrary and all members of the population face the same risks and rewards for interacting with in- and out-group members.}
\label{diagram.fig}
\end{figure}

\section*{Results}
To study polarization in a population faced with rising inequality or a declining economy, we apply methods from cultural evolution \citep{Kashima08,Boyd85,Cavalli81}.
This approach rests on the idea that each member of a large population employs a strategy that determines their propensity to interact with members of an in-group vs an out-group. We do not make any assumptions about the nature of these interactions other than that they provide differential benefits when successful and differential probabilities of failure (see below). Similarly, we do not make assumptions about the specific identity of in- and out-groups, rather we consider a simple base case where all sub-groups in a population find in-group interactions less risky but also less beneficial on average, than out-group interactions.

In this framework individual strategies are ``heritable'' via a copying process \citep{Traulsen:2006aa} in which individuals adopt the behavioral strategies of other members of the population with a probability that depends on the relative success of their respective strategies (see Methods). We assume that success is determined by a utility function, which depends non-linearly on the benefits received from individual interactions, as well as the state of the underlying economy (see Methods). We focus on a standard class of utility functions with an `S-shape' so that, if the benefits generated by a strategy become too low, there is a sharp decline in individual utility. This sharp decline can be thought of as an individual dropping below a poverty line, or an organization becoming insolvent. The full details and analysis of the model can be found in the Methods section below and in the SI (section 2-3).

We analyze the dynamics of polarization under two distinct sets of circumstances that have been identified previously as contributing to polarization \citep{McCarty16,Gidron}. First we consider a case in which the underlying economy starts to decline, pushing a large percentage of the population towards the poverty line. Second we consider the case of rising inequality in an otherwise stable economy.
\\
\\
\noindent\textbf{Economic decline:} In our model we distinguish between the {\em potential} and the {\em expected} benefit of interactions. The expected benefit is the probability that an interaction succeeds multiplied by the benefit it generates, i.e. $B_oq_o$ for out-group and $B_iq_i$ for in-group interactions. The potential benefit, by contrast, is simply the benefit received conditional on success, i.e. $B_o$ for out-group and $B_i$ for in-group interactions (Figure 1). We assume that out-group interactions always have greater potential benefit, $B_o>B_i$, but also lower probability of success, $q_o<q_i$. 

\begin{figure}[th!] \centering \includegraphics[scale=0.25]{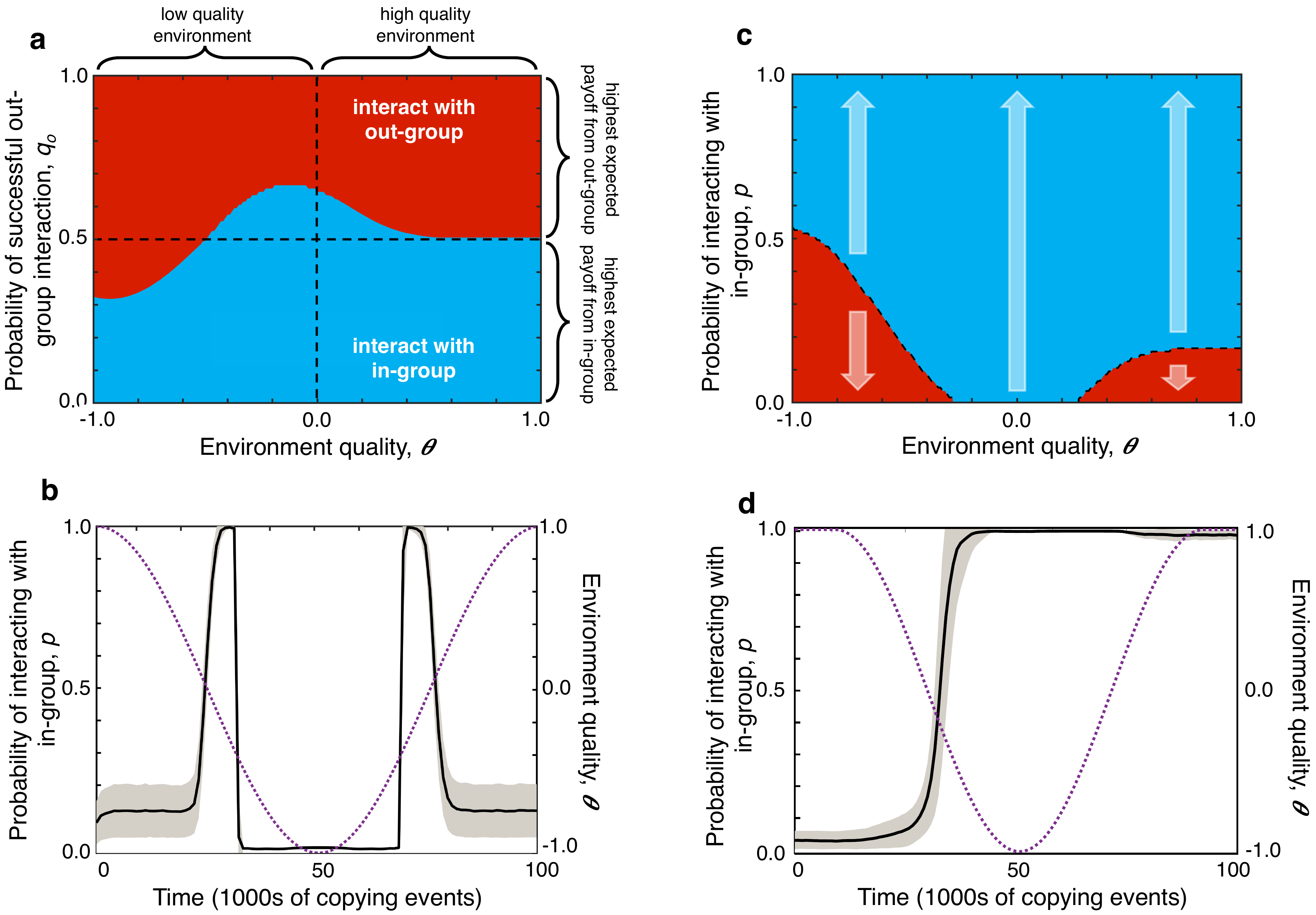}
\caption{\textbf{Polarization in a declining economy --} The evolutionary dynamics of polarization depend on the quality of the underlying economic environment, $\theta$, and the probability of successful out-group interactions, $q_o$. Here we fix the benefit for successful in- and out-group interactions at $B_i=0.5$ and $B_o=1$ and set the probability of successful in-group interactions to $q_i=1$. The quality of the economic environment $\theta$ varies from highly adverse ($\theta=-1$) to highly favorable ($\theta=1$). We calculate the strategy $p^*$ that maximizes utility from Eq. 4 (see Methods).  a) When the success of out-group interactions are independent of the strategy of the interaction partner, only one stable strategy evolves. Either a highly polarized ($p=1$, blue) or a highly diverse strategy ($p=0$, red). The stable strategy depends critically on the quality of the economic environment $\theta$ and is not reliably predicted by the expected payoffs from individual interactions. Here utility is assume to have non-linear sigmoidal component with threshold sharpness $h=10$, and a linear component with gradient of $\alpha=0.02$ (see  Methods Eq. 3 for full model details). The number of social interactions engaged in by individuals is $n=5$. Similar qualitative patterns are repeated for other choices of parameters and other choices of cumulative benefit function (see SI section 2). b) To illustrate the impact of these dynamics on a population in a changing economic environment, we carried out individual-based simulations where individuals copy more successful neighbors (selection strength $\sigma=10$, see Methods). The purple dotted line tracks the quality of the environment $\theta$, which varies sinusoidally whereas the black lines show the average population strategy.  The population size is fixed at $N=1000$ with success probabilities $q_i=1$ and $q_o=0.6$. Shown is the mean strategy at each point time (black line) across an ensemble of $1000$ populations, as well as the standard deviation of the strategy distribution for the ensemble at that time (gray region). Innovations, in which individuals try out novel strategies, occur at rate $\mu=0.001$ per copying event with size $\Delta=0.01$ (see Methods and SI Section 4 for full simulation details). c) When the success of out-group interactions depend on the strategy of the interaction partner, two or more strategies can be bi-stable in a given environment (see SI section 3). Arrows indicate the direction of evolution of strategy $p$ in a large population, in a given environment $\theta$. Blue regions indicate the basin of attraction for polarization $p^*=1$, while red regions indicate the basin of attraction for diversity $p^*=0$. Note that as the economic environment declines from favorable to unfavorable, diversity becomes unstable, while polarization is always stable. d) The evolutionary dynamics under individual-based simulations show how polarization increases in a declining economic environment, and remains even when the environment returns to being favorable. Parameters in c-d are the same as those given for a-b}
\label{results.fig}
\end{figure}

Even in cases where out-group interactions have higher expected payoff than in-group ones ($B_oq_o > B_iq_i$), there are circumstances in which it is better to behave in a risk-averse manner and to reduce risk by choosing in-group interactions. In a prosperous, high-quality economy, high-risk out-group interactions are favored whenever there is a greater expected payoff than that associated with in-group interactions, i.e. provided $B_oq_o > B_iq_i$ (Figure~\ref{results.fig}a). Thus high-quality economies support risk taking.

However, in a high-quality but declining environment, risk averse strategies become increasingly beneficial. Thus, there is a transition in the optimal behavioral strategy from out-group towards in-group interactions, i.e towards greater polarization (Figure~\ref{results.fig}a-b). Intuitively this transition occurs because, as the economy declines, failed interactions can result in a sharp decline in utility, pushing individuals toward insolvency. Thus transitioning to low-risk interactions becomes preferable. 

If the economic environment is very poor, by contrast, the inverse situation can arise, and risk-tolerant behavioral strategies can become optimal (Figure~\ref{results.fig}a lower-left quadrant). Intuitively this occurs because, under such an environment, rare successful interactions can push an individual back to solvency, and so ``gambling for redemption'' becomes the best strategy.



The results in Figure~\ref{results.fig}a-b describe a situation in which the success of in- and out-group interactions do not depend on the strategy of the interaction partner. As a result, the population is guaranteed to evolve towards a single strategy that maximizes individual payoff in that environment. However, the assumptions made here are not likely to hold in general --- often the success of an interaction depends on the strategies of both participants. 

In Figure~\ref{results.fig}c-d we therefore consider a case in which the success of out-group interactions depends on the strategies of both participants (see Methods). In this model an individual $i$ uses strategy $p_i$, i.e. she seeks an in-group interaction with probability $p_i$, and an out-group interaction with probability $1-p_i$. If $i$ seeks an out-group interaction with another individual $j$, we assume that the interaction succeeds with probability $q_o(1-p_j)$, where $p_j$ is the strategy of $j$. Thus if $j$ is only willing to engage in in-group interactions, i.e. $p_j=1$, the out-group interaction with $i$ will surely fail. 


In this case the system becomes bi-stable in a high quality environment, with both a high polarization and a low polarization state possible (see SI section 3). However, as environmental quality declines, the low polarization equilibrium is frequently lost (Figure~\ref{results.fig}c), and a low polarization population tends to evolve rapidly towards the high polarization state (Figure~\ref{results.fig}d). Crucially however, the converse does not occur for a population in a high-polarization state in an improving environment. Rather, the high polarization state is always stable. The low polarization state, once lost, is thus hard to recover via a process of cultural evolution. This remains true even when the low polarization state would produce higher individual utility for all members of the population. Polarization, under this model, takes on the character of a social dilemma, in which there would be a collective benefit if everyone switched to a low polarization strategy, but each individual has incentives to maintain a high polarization strategy. Recovering low polarization would therefore require 
coordination, possibly introduced in response to an exogenous shock altering salience of the polarized identities relative to the identity of the full population as an aggregate.

Note that, just as previously, a risk-tolerant ``gambling for redemption'' strategy becomes stable in a very poor environment (Figure~\ref{results.fig}c). However the system remains bi-stable, and so loss of polarization is not inevitable even under these circumstances. 

\clearpage

\noindent\textbf{Rising inequality:} We have shown that economic decline can facilitate the emergence of polarization by inducing individuals to switch to more risk-averse strategies. Under rising inequality, even in an overall favorable environment, the less well-off subset of the population can face similar incentives. 

We explored the effect of such inequality on the evolutionary dynamics of polarization by assuming that a proportion $\pi$ of the population are very well-off while the remaining $1-\pi$ are less well-off. We assume that the underlying economic environment experienced by the more well-off is $\theta_+=\theta_0+\frac{1-\pi}{\pi}\theta_g$ while the environment for the less well-off is $\theta_-=\theta_0-\theta_g$, so that the average environment is $\theta_0$ and the Gini coefficient for the population is $g=\frac{\theta_g}{\theta_0}(1-\pi)$.

\begin{figure}[th!] \centering \includegraphics[scale=.19]{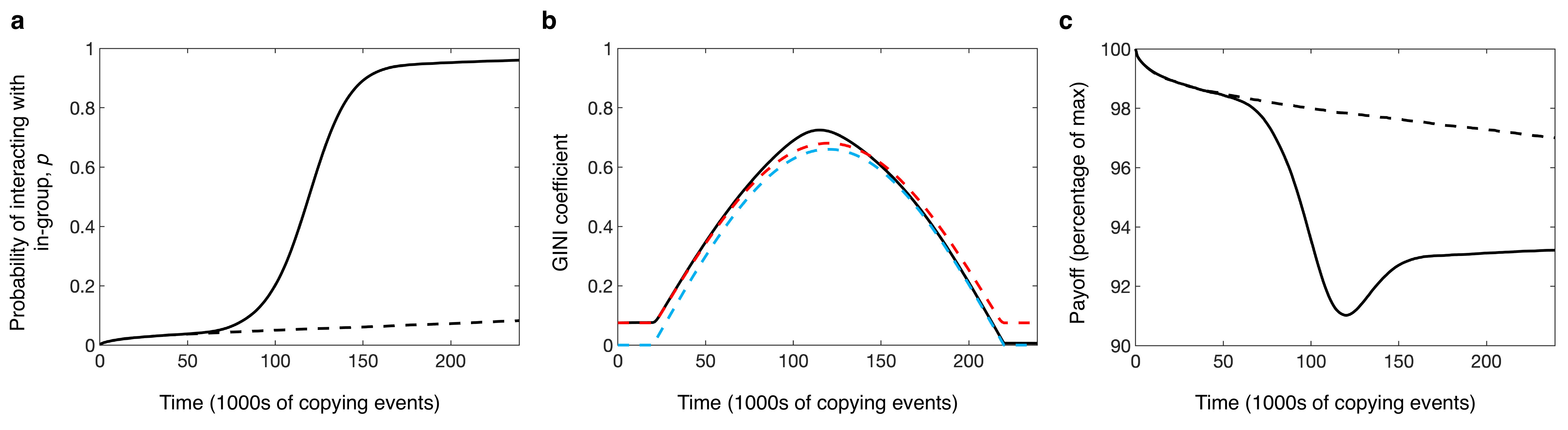}
\caption{\textbf{Polarization under rising inequality --} We ran individual-based simulations in an economic environment with evolving levels of inequality. Here we assume sinusoidally varying inequality with $\theta_0=1.5$ and $\theta_g\in[0,1.1]$ and $\pi=0.1$. All other parameters are those provided in Figure~\ref{results.fig}c-d  a) Under rising inequality, polarization sharply increases (solid line) compared to the case with no inequality (dashed line). b) The Gini coefficient for the population (solid black line) changes as the inequality in the underlying economic environment, $\theta_g$ changes, and as behavioral strategies evolve. Also shown for comparison are the Gini coefficients for a population using a fixed strategy of either polarization ($p=1$, blue dashed line) or diversity ($p=0$, dashed red line). c) The average benefits from social interactions decline under rising inequality (solid line) and do not recover to the same levels as when inequality was absent (dashed line), even when the environment is no longer unequal.
} 
\label{model2.fig}
\end{figure}

In the case where success of out-group interactions depends on the strategy of the interaction partner, as in Figure~\ref{results.fig}c-d, we find that an increase in inequality has a similar effect to a decline in the economic environment (Figure~\ref{model2.fig}a). As the environment of the less wealthy declines (Figure~\ref{model2.fig}b) they adopt risk averse strategies, which rapidly spread to the whole population, resulting in high levels of polarization. However the situation is not reversed when inequality declines, due to the bi-stability of the system. Once again, we see that this irreversible polarization has the character of a social dilemma, with the average benefits gained by the population from social interactions remaining stuck in a sub-optimal state (Figure~\ref{model2.fig}c).
\\

\clearpage

\noindent\textbf{Affective polarization in the United States:} 
Recent research in political science has stressed that partisanship is a salient social identity as well as a marker for the various other group identities that have become associated with the major parties
\citep{Iyengar:2012,Iyengar:2015,Mason:2015,Mason:2018}.  A useful measure of group-based partisanship known as affective polarization is measured by the difference in ``warmth'' towards the preferred and non-preferred major US political party (i.e. Republican or Democrat), reported by individuals via a feeling thermometer scale \citep{ANES,DVN/ADYZFU_2019,DVN/0UAZ5C_2019,DVN/HCBA4U_2019}. Given its association with identity-based politics, affective polarization is a measure well-suited to evaluating the predictions of our group-based model, that inequality and inter-group conflict are mutually causal and therefore correlated. We evaluate the empirical case for an association between inequality and affective polarization in the United States. Recent analysis has shown that inequality and affective polarization are correlated over time in a panel of developed democracies \citep{Gidron}. Here we employ a similar approach to look at the correlation across states within the US over the course of three presidential election cycles from 2008--2016, using publicly available ANES and CCES survey data \citep{ANES,DVN/ADYZFU_2019,DVN/0UAZ5C_2019,DVN/HCBA4U_2019}.

\begin{figure}[th] \centering \includegraphics[scale=0.25]{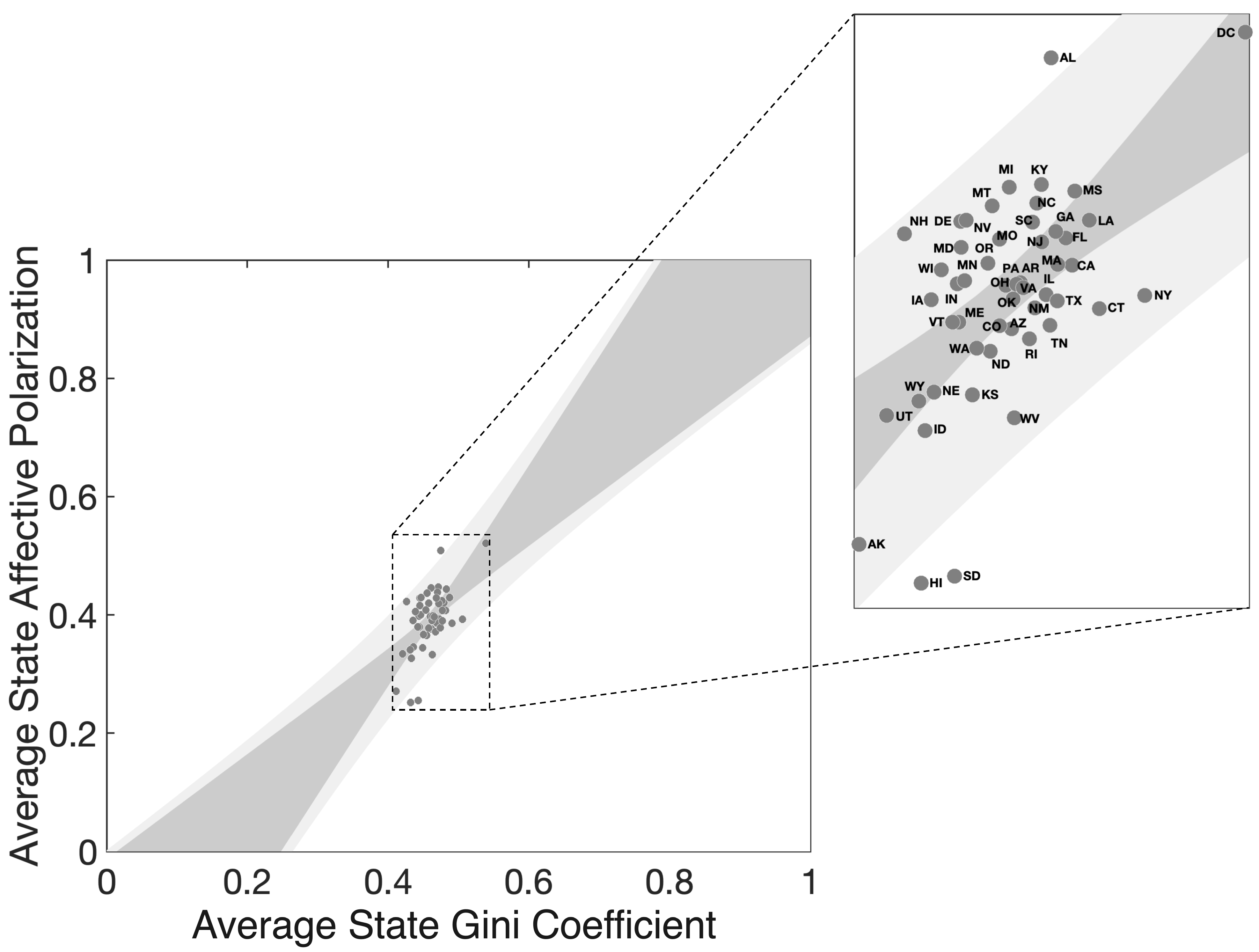}
\caption{\textbf{Affective polarization and inequality in the US --} We show the correlation for the pooled data across all three election cycles between state-level affective polarization estimated from \citep{ANES,DVN/ADYZFU_2019,DVN/0UAZ5C_2019,DVN/HCBA4U_2019} and state-level Gini coefficient taken from \citep{Census}. The dark gray region gives the $95\%$ confidence interval and the light gray region the $95\%$ prediction interval for the model. The expanded region shows individual state-level vales. The full data set and additional analyses can be found in SI section 5.} 
\label{ANES.fig}
\end{figure}

 Figure~\ref{ANES.fig} shows the positive  correlation (significant with $p<0.01$, two-tailed t-test, $t=5.2$) between state-level Gini \citep{Census} and the state-average affective polarization, in the pooled data across all three election cycles. A two-way fixed-effects model with election-specific intercepts gives similar results (significant positive correlation with $p<0.01$, two-tailed t-test, $t=4.4$, see SI section 5). Additional robustness checks controlling for demographic factors, and individual level regressions are presented in SI section 5, and yield similar results.

\clearpage
 \section*{Discussion}


Having provided an account of polarization connecting economic well-being, inequality, and group conflict, we now discuss how this account may generalize, and how it provides explanations and clarity for several salient questions in  political economy.




First, our model provides a micro-foundation for the links between intra-group cooperation and conflict and economic hardship.  Our argument is most closely related to the well-known model of Realistic Conflict Theory (RCT) which postulates that group behavior and prejudice is based on material incentives. RCT postulates that economic hardship can induce greater inter-group hostility by enhancing competition over scarce resources. Thus, it predicts a correlation between aggregate economic output and group-level cooperation, an effect our models exhibit in the main, although we also show cooperation can emerge in some states of extreme as a consequence of `gambling for redemption'.  As that indicates, the mechanisms underlying our models are quite different.  Instead of postulating direct conflict over scarce resources as per RCT, our model suggests that during economically-challenging periods, individuals may simply find it too risky to cooperate with the out-group. Of course, in situations of extreme deprivation such circumstances may set the stage for between-group competition for resources.  Thus, our model is congruent with findings that diversity only reduces public goods investment when countries suffer economic stress \citep{Wimmer16}. 
Moreover, RCT has also been criticized for not providing adequate explanation or motivation for in-group biases.  Our framework provides such an explanation based on the idea that within-group transactions are less risky than cross-group transactions.

Finally and perhaps most importantly, our theory argues that economic stagnation and group conflicts can be mutually causal.  When people withdraw from the more profitable, but riskier, out-group transactions, both aggregate and {\em per capita} output necessarily fall. This has a self-reinforcing effect as the fall in economic output engenders even lower levels of out-group interaction. These positive feedback loops are similar to those describing interactions between ideological polarization and economic inequality \citep{McCarty16}.  


In extreme cases economic shocks may even lead to civil and social conflict \citep{Rodrik:1999, Miguel:2004}.  Chassang and Padro-i Miquel\citep{Chassang:2009} develop a formal model of the causal impact of economic shocks on civil war onset.  Their argument is that shocks produce conflict because they reduce the opportunity cost of fighting in the short run.  Our argument in contrast focuses on how economic stress makes intra-group cooperation far more risky.  Without the offsetting benefits of such cooperation, conflicts are more likely to escalate. 

Returning to our motivating example of the recent electoral success of nationalistic candidates and movements, our theory provides an account of the impact of economic performance on political preferences and identities. Consistent with our model, Dorn et al. \citep{Dorn:2016} recently showed that the legislative representatives of regions hit hard by the surge in US imports from China have elected more right-wing representatives.  Similarly, other authors \citep{Funke:2016,Mian:2014} have shown that calamitous economic shocks such as the Great Depression and the Global Financial Crisis increased support for right-wing politicians, especially those of the far right who denigrate social out-groups. Economic stress and fear 
have been shown to be excellent highly-localized predictors 
for the success of a number of populist movements including those in the Ukraine\citep{Zhukov16} as well as the United States and United Kingdom\citep{Obschonka18,Becker17}.


Our model also makes predictions that relate polarization to wealth inequality. For the majority of the population, an increase in inequality involves a steep decline in relative well-being compared to the most affluent. In our model, inequality drives polarization because the risk aversion of the less well-off spreads to the whole population.  However it is also possible that perceptual changes due to a change in relative affluence may be sufficient to make individuals risk averse and withdraw from out-group interactions, leading to a self-fulfilling cycle of economic decline. 

 
Our modeling framework may also help explain the observed causal impact of  income inequality on the rise of political conservatism \citep{McCartyShor16}.  While previous models highlight a tendency for inequality to empower the left due to increased demands for redistribution \citep{Meltzer81}, our framework suggests a more general withdrawal from out-group interaction in such contexts.  Such an orientation towards the in-group over the out-group is often associated with the political right \citep{Haidt12}. Similarly, Hogg \citep{Hogg14} has proposed that when individual identity is weak --- a situation possibly triggered by loss of relative status --- there is an increased tendency to take identity from a group instead, and to simultaneously prefer extremism, strong leaders, and the inhibition of dissent. While leadership is not a feature of our framework, the increased in-group preference we observe under economic decline would be facilitated by a capacity to strongly signal in-group membership to potential collaborators. Expressing allegiance to leaders proclaiming bold and ordinarily socially-costly opinions may be one mechanism for such signaling.

Perhaps the biggest take-away from our model is a prescriptive one, for how to overcome polarization once introduced. Our work highlights an asymmetry in how easy it can be to move by gradual change from low to high polarization, and back. The persistence of a high-polarization equilibrium in our model is caused by a coordination problem, where most individuals would benefit from diverse interactions but are unable to discover a means for simultaneous collective exchange. An external event may be needed to provide such coordination. During significant national crises such as wars, states have very strong incentives to solve these problems.  
Thus exogenous events that (re)establish the importance of larger-scale (e.g. national) identities may be required to regenerate inter-group cooperation and reduce polarization.  This argument is supported by a review of twenty studies,\citep{Bauer16} which show radically increased altruism in populations after experiencing war. 
Recent work \citep{Scheidel17} has also shown that external shocks such as war are a necessary condition for reducing inequality.
At the level of public policy, Scheve and Stasavage \citep{scheve2016taxing}  show that progressive taxation is largely the product of fairness norms forged during wartime. Jong et al. \citep{Jong15} show that other less devastating negative circumstances such as terrorist attacks can also lead to identity fusion.

Such findings may also relate to our observation that when the economy is {\em extremely} weak, working with out-group members can again become the best strategy. If recovery after events such as war, drought, public health crisis or embargo is sufficiently rapid, this may bypass a dip back into parochialism and leave a society less polarized.


Prescriptively then, if political parties and governments wish to tackle high levels of inter-group conflict and low levels of trust, they should consider the underlying economic basis of these conflicts.  In particular, social arrangements should be devised to adequately support the population at a level that makes out-group interaction remain viable.  And to the extent that individual out-group interactions are risky, society may wish to spread those risks through public means. Periods of crisis may well provide the opportunity to not only create such economic solutions, but to coordinate public opinion and reestablish a larger, more inclusive sense of identity.



\section*{Methods}

In order to capture variation in behavioural diversity and its consequences for polarization we adopt a model derived from the study of cultural evolution and evolutionary game theory, under which individuals accumulate benefits through multiple interactions with other members of a finite population of $N$ individuals.

We assume that each individual is faced repeatedly with the choice of interacting either with someone who is ``like'' them (in-group interactions) or ``unlike'' them (out-group interactions), where in-group interactions provide a benefit $B_i$ with a probability $q_i$ and benefit $0$ with probability $1-q_i$ (Figure~\ref{diagram.fig}). That is, the risk of failure in an in-group interaction is $1-q_i$, and we assume here that there is no further penalty for failure than missed opportunity. Increasing the cost of failure does not qualitatively affect our outcomes (see SI). Similarly, an out-group interaction provides a benefit $B_o$ with probability $q_o$ and benefit $0$ with probability $1-q_o$. As discussed in the introduction, we make the key assumption that out-group interactions come with higher reward ($B_o>B_i$) but also higher risk ($p_o<q_o$).

Each individual is assumed to participate in $n$ interactions, whose success or failure goes to determine the total payoff accumulated by the individual during that time period, the where the number of available in- and out-group interactions is assumed very large and consequently $N{\gg}n$. Typically we assume $n<10$, reflecting an individual who is making a decision based on a few sources of information. We discuss the case of larger numbers of in- and out-group interactions $n$ in the SI. Each individual is then characterized by a strategy $p$ which gives the probability that they choose an in-group interaction, and consequently each individual chooses an out-group interaction with probability $1-p$ (Figure~\ref{diagram.fig}).

Given this model, the probability that a player with strategy $p$ engages in $l_i$ successful in-group interactions out of a total $k$ in-group interactions and $l_o$ successful out-group interactions out of a total $n-k$ out-group interactions is given by

\begin{equation}
\pi(k,l_i,l_o | n)={n\choose k}p^k(1-p)^{n-k}{k\choose l_i}q_i^{l_i}(1-q_i)^{k-l_i}{n-k\choose l_o}q_o^{l_o}(1-q_o)^{n-k-l_o}
\end{equation}
\\
that is, the number of in- and out-group interactions and the number of successful interactions each follow binomial distributions. The resulting expected benefit derived form  successful interactions under this model is then simply 

\begin{equation}
\sum_{k=0}^n\sum_{l_i=0}^k\sum_{l_o=0}^{n-k}\pi(k,l_i,l_o|n)(B_il_i+B_ol_o)=nB_iq_ip+nB_oq_o(1-p)
\end{equation}
\\
and the strategy that maximizes Eq. 2 is either $p=1$ (always interact with in-group) if $B_iq_i>B_oq_o$ and $p=0$ (always interact with out-group) otherwise.
However such a linear model does not in general reflect the reality of the way benefits accumulate either biologically, in an ecosystem or in human society. In many situations a minimum level of resources is required to achieve a particular goal (e.g. avoid starvation or reproduce in a biological system; purchase property or start a business in an economy).  Income above that threshold, while still advantageous, is less beneficial. That is, benefits tend to accumulate non-linearly.

At the same time, both ecosystems and economies may be influenced by exogenous factors (such as weather events) so that they expand or contract the per-capita resources available for a population of given size. When such fluctuations occur, the non-linear accumulation of benefits described above leads to changes in the curvature of the utility function of a given individual, and thus their level of risk aversion. Since in- and out-group interactions differ both in their level of expected benefit and their level of risk, this leads to changes in behavior. We consider the evolutionary dynamics of behavior both in the case where the risk of out-group interactions is fixed $1-q_o$, and where it depends on the ``willingness'' of out-group members to engage in such interactions i.e. where the risk associated with out-group interactions depends on the strategy adopted by other members of the population. 

To understand the consequences of shifting environments and non-linearly accumulating benefits on individual behavior in our model, we consider the evolutionary dynamics of the system. We consider a population evolving under a ``copying process'' \citep{Traulsen:2006aa} in which individuals are able to observe the ``fitness'' -- i.e. the total benefit accumulated via in- and out-group interactions -- of other individuals and compare it to their own. The dynamics of the model are as follows: An individual $h$ is chosen at random from a population of fixed size $N$. A second individual $g$ is then chosen at random for her to ``observe''. If $h$ has fitness $w_h$ and $g$ has fitness $w_g$ then $h$ chooses to copy the strategy of $g$ with probability $1/(1+\exp[\sigma(w_g-w_h)])$, where $\sigma$ scales the ``strength of selection'' of the evolutionary process. Note that if $w_g\gg w_h$ the probability of $h$ copying the behavior of $g$ is close to 1, whereas if $w_g\ll w_h$ the probability is close to 0. 

In order to explore the evolutionary dynamics of the system we must also specify how fitness $w$ depends on the benefits received from individual in- and out-group interactions, $B_i$ and $B_o$. In order to model the non-linear accumulation of ``fitness''' benefits from diverse social interactions across a range of environments, we assume that the linear accumulation of fitness benefits is modified by a sigmoidal function, such that

\begin{equation}
w(k,l_i,l_o | n)=\frac{\exp[h(l_iB_i+l_oB_o+n\theta)/n]}{1+\exp[h(l_iB_i+l_oB_o+n\theta)/n]}(1+\alpha(l_iB_i+l_oB_o+n\theta))
\end{equation}
\\
where $h$ controls the ``steepness'' of the sigmoid (how sensitive fitness is to changes in accumulated benefits), $\alpha$ controls the rate of linear accumulation of benefits and $\theta$ controls the environment, so that when $\theta$ is large (relative to accumulated benefits) and positive, the sigmoidal term is close to 1 and fitness tends to accumulate linearly. Conversely when $\theta$ is large and negative (relative to accumulated benefits) fitness tends to be close to 0. The form of Eq. 3 reflects an environment in which a certain minimum level of benefit is required for success or survival.

From Eq. 3 we can calculate the expected fitness $\hat{w}$ of a player with strategy $p$, under the model with fixed risk, which is simply

\begin{equation}
\hat{w}=\sum_{k=0}^n\sum_{l_{i}=0}^k\sum_{l_o=0}^{n-k}\pi(k,l_i,l_o|n)\frac{\exp[h(l_iB_i+l_oB_o+n\theta)/n]}{1+\exp[h(l_iB_i+l_oB_o+n\theta)/n]}(1+\alpha(l_iB_i+l_oB_o+n\theta))
\end{equation}
\\
In order to characterize the evolutionary dynamics of this system, we use Eq. 4 to determine how the strategy $p^*$ that maximizes Eq. 4 varies with the environment, $\theta$, and the probability of success in interactions with in- and out-group members, $q_i$ and $q_o$.  Since Eq. 4 cannot be treated analytically in general we calculated numerically the strategy $p^*$ that will maximize fitness as a function of the environment and the probability of successful in- and out-group interactions, and show that for a given environment and risk level, there is a single global optimal strategy for the system (see Figure~2a-b and SI).

Finally, we consider a version of our model that includes the possibility that the success of out-group interactions depends on the strategy adopted by the out-group player. We assume for simplicity that in-group members are always willing to interact -- however we relax this assumption in the SI. We then assume that a successful out-group interaction between two players $g$ and $h$ depends on both players'  willingness to interact, i.e. on $p_g$ and $p_h$. That is, we set $q^{gh}_o=q_o(1-p_h)$ where $q_o$ is the intrinsic probability of success and $(1-p_h)$ is the probability that player $h$ agrees to participate in the interaction.  In order to explore the evolutionary dynamics of this system we adopt the framework of adaptive dynamics \citep{Geritz98,Doebeli:2004aa} to calculate the stable strategies of the model under small changes to a player's strategy $p$. The fitness of a strategy $p_h$ in a population of players using a resident strategy $p$ is 

\begin{align}
\nonumber\hat{w}_h=\sum_{k=0}^n\sum_{l_{i}=0}^k\sum_{l_o=0}^{n-k}{n\choose k}p_h^{k}(1-p_h)^{n-k}\times\\
\nonumber {k\choose l_i}q_i^{l_i}(1-q_i)^{k-l_i}{n-k\choose l_o}(q_o(1-p))^{l_o}(1-q_o(1-p))^{n-k-l_o}\times\\
\frac{\exp[h(l_iB_i+l_oB_o+n\theta)/n]}{1+\exp[h(l_iB_i+l_oB_o+n\theta)/n]}(1+\alpha(l_iB_i+l_oB_o+n\theta))
\end{align}
\\
and we can calculate the stability of the resident strategy $p$ to invasion by calculating the selection gradient

\begin{equation}
s=\frac{\partial \hat{w}_h}{\partial p_h}\Bigg|_{p_g=h}
\end{equation}
\\
which determines the local evolutionary dynamics of the system. Once again, we explore the equilibria of the system numerically, and show that the system is frequently bi-stable (Figure~2c-d), and for some parameter choices has three stable equilibria (see SI).
\\
\\
\textbf{Invasion:}
We consider the evolutionary dynamics under the copying process as described in the main text \citep{Traulsen:2006aa}, under which the probability that a player with strategy $g$ copies the strategy of another player $h$ is

\begin{equation}
r_{g,h}=\frac{1}{(1+\exp[\sigma(w_g-w_h)])}
\end{equation}
\\
and the resulting growth rate of a rare mutant $h$ in a population with resident strategy $g$ is

\begin{equation}
S(h,g)=\frac{r_{g,h}}{r_{h,g}}=\frac{(1+\exp[-\sigma(w_g-w_h)])}{(1+\exp[\sigma(w_g-w_h)])}=\exp[-\sigma(w_g-w_h)]
\end{equation}
\\
Switching without loss of generality to log-fitness (and ignoring the proportionality constant) we can then simply write

\begin{equation}
s(h,g)=w_h-w_g
\end{equation}
\\
Where if $s>0$, $h$ is increasing in frequency.
In order to construct pair-wise invasibility plots (see SI section 2-3) we then simply look at the sign of Eq. 8-9 when $w$ is given by Eq. 3-5. Note that in the first case we analyze (Eq. 3-4) the payoff $w$ depends only on the focal player's strategy (i.e. the fitness of the resident and the mutant do not depend on one another). This case is formally similar to an optimal foraging model with a sigmoidal functional response curve.

A strategy $h=g=g^*$ is a local ESS if and only if

\begin{equation}
\frac{\partial^2 s(h,g)}{\partial h^2}<0
\end{equation}
\\
when evaluated at $g^*$, which must be a point of zero selection gradient. The strategy  $g^*$ is convergence stable if and only if \citep{Geritz98}

\begin{equation}
\frac{\partial^2 s(h,g)}{\partial g^2}>\frac{\partial^2 s(h,g)}{\partial h^2}
\end{equation}
\\
 when evaluated at  $h=g=g^*$. We use Eqs. 9-11 in constructing invasibility plots and determining the character of singular points (see SI section 2-3).
 
\newpage

\section*{Supplementary Information}

\noindent In this supplement we describe relaxations to the modeling assumptions presented in the main text to demonstrate the robustness of our conclusions. In particular we vary the ``functional response curves'' that relate the outcome of a given social interaction to the accumulated utility of many interactions and the number of interactions among individuals. We also provide further details of the individual-based simulations presented in the main text. Note that equation numbers continue from the main text.
\setcounter{equation}{11}

\section{Model Parameters}

The parameters associated with the model presented in Figure 1 and the main text are summarized in Table 1 below. We now discuss how variation in the parameters of the model impacts the associated evolutionary dynamics and the degree of polarization that arises across environments. We first discuss the case where the probability of success of an out-group interaction is constant and independent of the strategies adopted by other members of the population. We then discuss the case where the risk of out-group interactions depends on the strategy of the target for the interaction. We analyze both models by looking at the equilibria under both local and non-local mutations (i.e under scenarios where players adjust their behavior either gradually or in sudden-large shifts such as may occur in response to structural change).

\begin{table}[h]
\begin{center}
    \begin{tabular}{ | l | l | p{9cm} |}
    \hline
    Parameter & Default Simulation Value & Meaning \\ \hline
    $B_i$ & 0.5  & Benefit received due to a successful in-group interaction. \\ \hline
    $B_o$ & 1.0 & Benefit received due to a successful out-group interaction. \\ \hline
   $q_i$ &  1.0 & Probability of a successful in-group interaction \\ \hline
   $q_o$ &  0.6 & Probability of a successful out-group interaction \\ \hline
   $n$ &  5 & Number of attempted interactions before strategy update \\ \hline
   $N$ &  1000 & Population size \\ \hline
   $\theta$ &  $[-1,1]$ & Quality of the environment (no inequality) \\ \hline
   $\theta_0$ &  $[0,1]$ & Average quality of the environment (with inequality) \\ \hline
    $\theta_g$ &  $[0,1.1]$ & Strength of inequality \\ \hline
    $\pi$ &  $0.1$ & Proportion of the population that is very well off \\ \hline
   $h$ &  2 & Steepness of the functional response curve \\ \hline
   $r$ &  0.01 & Slope of the functional response curve \\ \hline
   $\sigma$ &  10 & Selection strength \\ \hline
   $\mu$ &  0.0001 & Mutation rate \\
    \hline
    \end{tabular}
        \caption{\small Model parameters and the default values chosen for main text individual-based simulations}
\end{center}
\end{table}

\clearpage

\section{Case 1: Non-social payoffs}

\subsection{Stability and Invasibility}

We first consider the case in which the probability of success of an out-group interaction is simply $q_o$, which does not depend on the strategy of the target of the interaction. Thus the payoff $w_h$ of a mutant $h$ does not depend on the background $g$ into witch it is introduced. This means that a strategy that maximizes $w$ can always invade and can never be invaded, so that under an evolutionary process with non-local mutations as shown in Figure 2 (main text) the population will always arrive at the global maximum.

However, we are also interested in the behavior of the system under local mutations (or ``gradual methods'' as they are called in the main text). To this end we look at the selection gradient of main text Eq. 4 which gives

\begin{align}
\nonumber \frac{\partial s(f,g)}{\partial f}=\sum_{k=0}^n\sum_{l_{i}=0}^k\sum_{l_o=0}^{n-k}{n\choose k}
p_f^{k-1}(1-p_f)^{n-k-1}(k-np_f)\times\\
\nonumber {k\choose l_i}q_i^{l_i}(1-q_i)^{k-l_i}{n-k\choose l_o}(q_o)^{l_o}(1-q_o)^{n-k-l_o}\times\\
\frac{\exp[h(l_iB_i+l_oB_o+n\theta)]}{1+\exp[h(l_iB_i+l_oB_o+n\theta)]}(1+r(l_iB_i+l_oB_o))
\end{align}
\\
which we can evaluate numerically to calculate the points of zero selection gradient as shown in Figure S1 below. We can also use Eq. 12 to gain insight into the dynamics of polarization on display in Figure 2 of the main text.

In particular, if the environment is sufficiently bad that $l_iB_i+l_oB_o+n\theta<0 \ \ \forall \ l_i, l_o$ or if it sufficiently good such that $l_iB_i+l_oB_o+n\theta>0, \ \ \forall \ l_i, l_o$ then we can approximate the sigmoidal term in Eq. 12 as a constant and recover selection gradient

\begin{align}
\nonumber \frac{\partial s(h,g)}{\partial f}=\sum_{k=0}^n{n\choose k}
p_f^{k-1}(1-p_f)^{n-k-1}(k-np_f)(1+nrq_oB_o+kr(q_iB_i-q_oB_o))\\
=r(q_iB_i-q_oB_o)
\end{align}
\\
i.e. evolution will proceed in the direction of the the strategy that increases expected fitness. However, for intermediate values of $\theta$ we can approximate the sigmoidal term as 0 for $l_iB_i+l_oB_o<-n\theta$ and as 1 otherwise. Thus Eq. 12 becomes the sum over the probability distribution conditional on the fact that the payoff received is greater than $-n\theta$. This cannot be calculated explicitly in most cases but note that if $nq_iB_i+nq_oB_o+n\theta>0$ then terms with low values of $l_o$ or $l_i$ will be eliminated from the summation. Since $q_o<q_i$, this means that players who tend to use out-group interactions will tend to suffer more in this regime, and thus the population becomes risk averse. In contrast, when $nq_iB_i+nq_oB_o+n\theta<0$ only terms with high values of $l_o$ or $l_i$ will be included in the summation, which tends to favor out-group interactions. This qualitatively captures the results shown in Figure 2. 

In the following sections we systematically vary the parameters of Table 2 in order to assess the robustness of the results presented in the main text.

\begin{figure*}[h] \centering \includegraphics[scale=0.2]{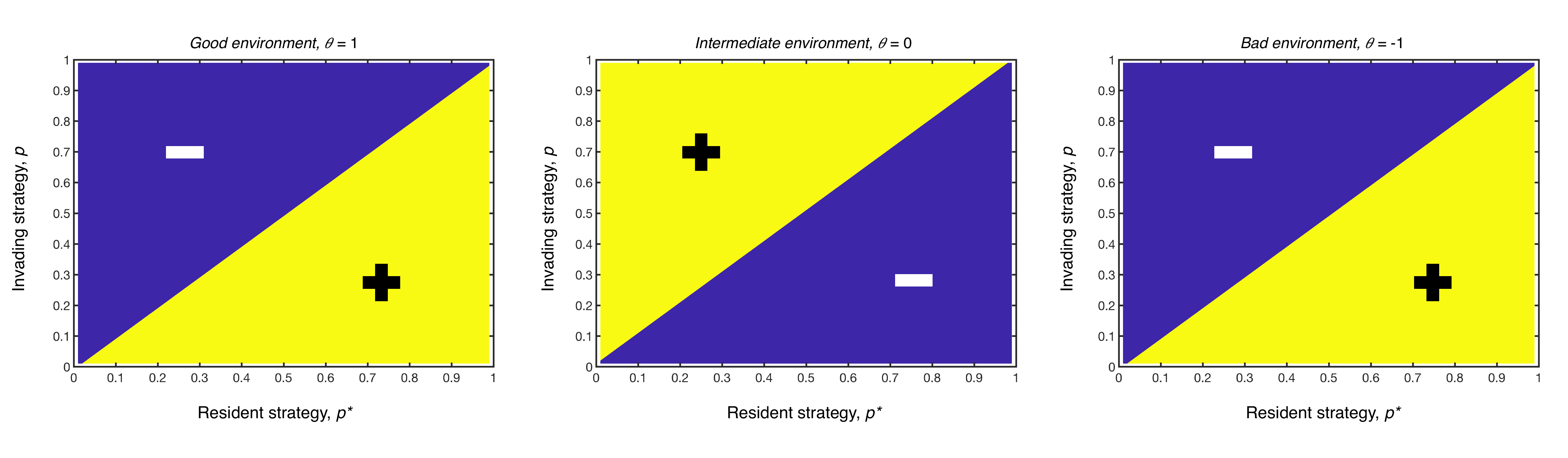}
\caption*{\small Figure S1 --  Pairwise invasability plot in different environments using the default parameters as given in Table 1. We see, just as in Figure 2 of the main text, that low polarization ($p^*=0$) is stable in a very good or a very bad environment (left and right plots) but that this situation is reversed in an intermediate environment (center plot). Under both local and non-local mutations this effect is evident}
\end{figure*}

\clearpage

\subsection{Impact of Risks and Benefits of Interactions}

Figure S2 shows the strategies that maximize fitness across environments as we vary the risk parameters $q_i$ and $q_o$ and the benefit parameters $B_i$ and $B_o$. In all cases we see qualitatively similar results to those shown in Figure 2 of the main text -- for intermediate environments risk aversion can lead to an increase in polarization even when the expected benefit of out-group interactions exceed those of in-group interactions.

\begin{figure*}[h] \centering \includegraphics[scale=0.27]{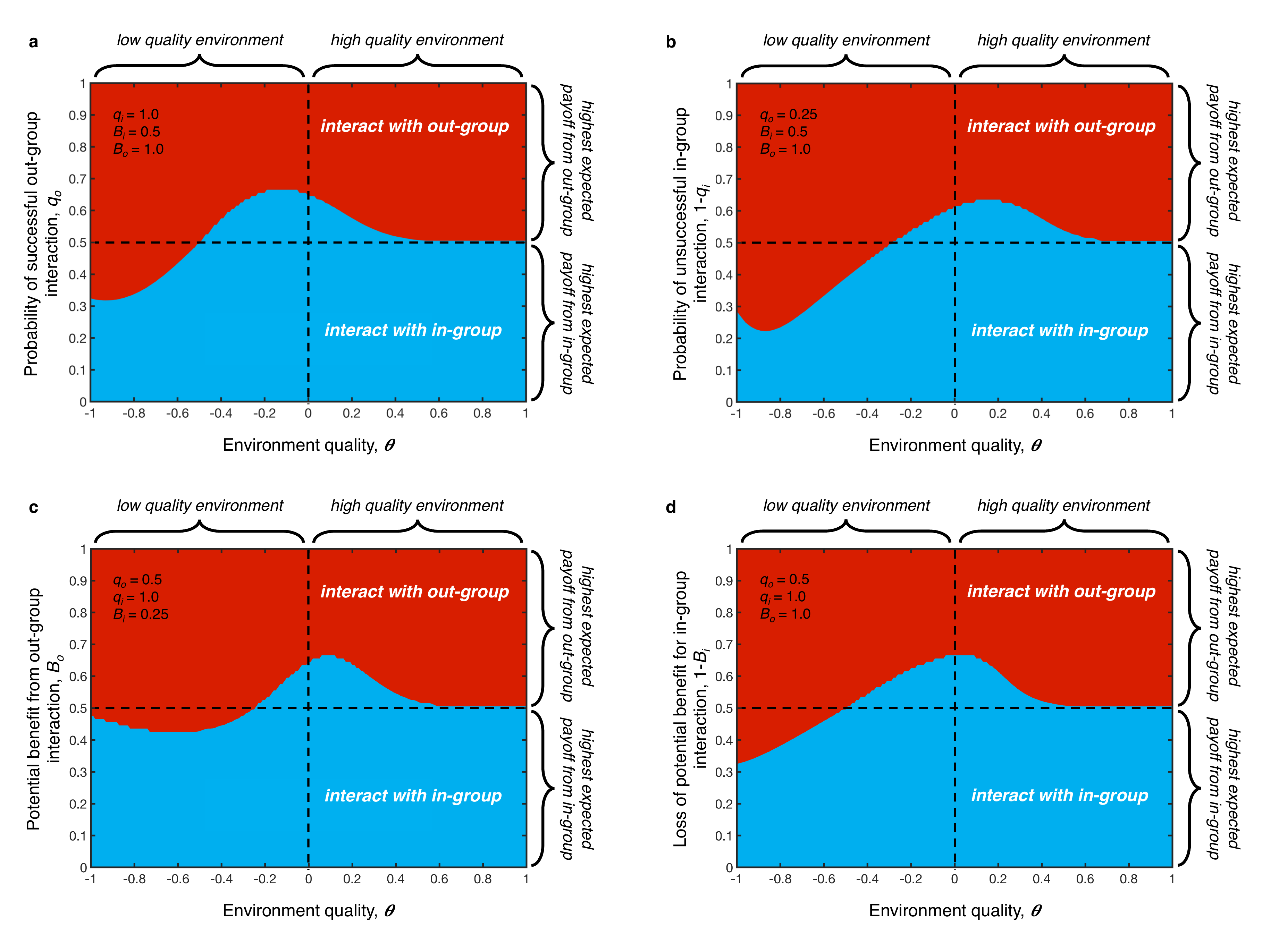}
\caption*{\small Figure S2 -- Stable equilibria as a function of environment ($\theta$, x-axis) and a) The probability of success of out-group interactions, b) the probability of failure of in-group interactions, c) the benefit of successful in-group interactions and d) the opportunity cost of a successful in-group interaction compared to a successful out-group interaction. All other parameters are set to the default values in Table 1. }
\end{figure*}
\clearpage

\noindent We also examined the stable strategies of the model fixing the expected benefits of in- and out-group interactions $B_iq_i$ and $B_oq_o$ and varying the risk associated with out-group interactions. Once again we see a shift from stable low-polarization strategies at intermediate environments, unless the risk of out-group interactions becomes low (i.e $q_o$ becomes sufficiently large) in which case low polarization strategies are always stable.

\begin{figure*}[h] \centering \includegraphics[scale=0.27]{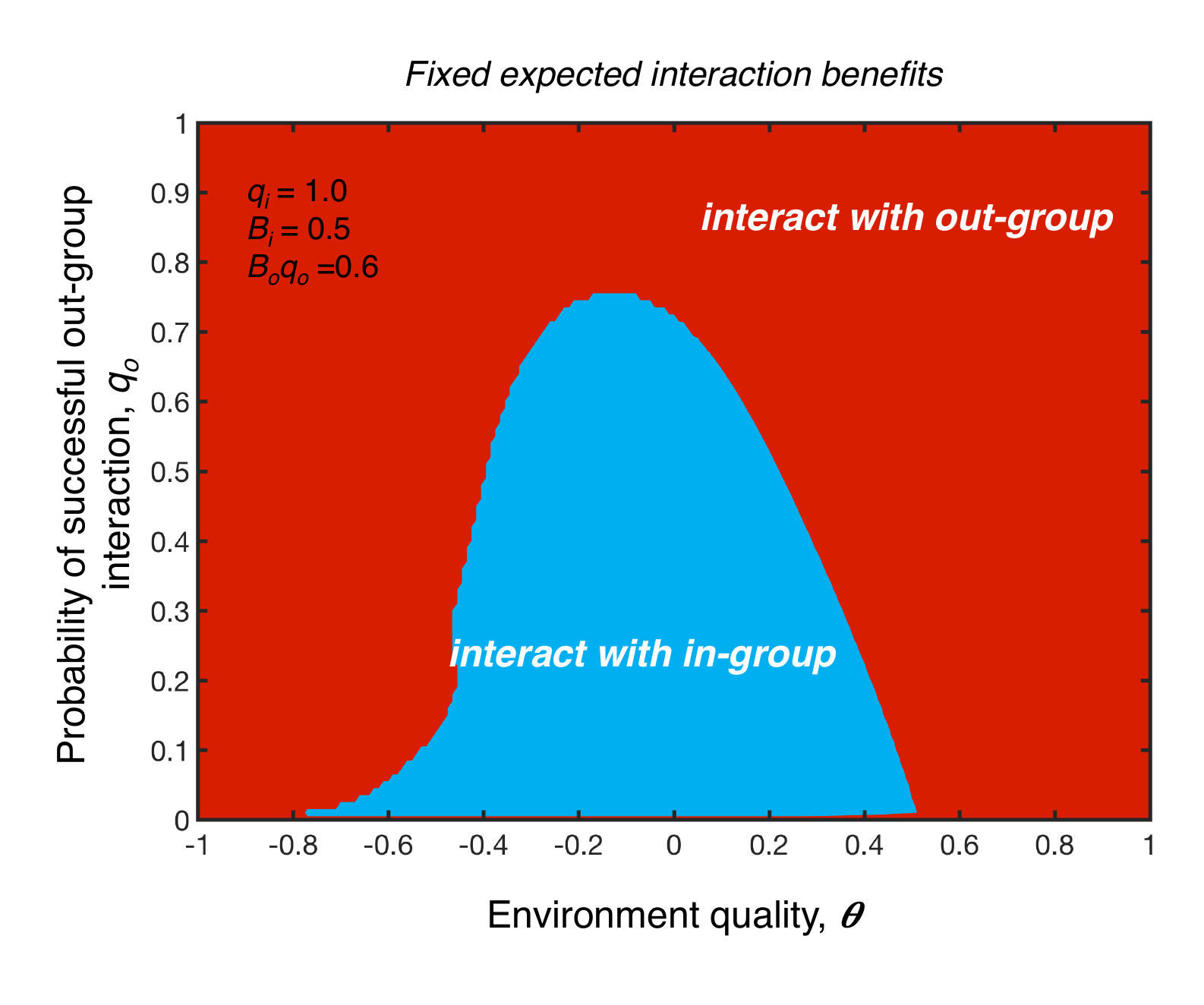}
\caption*{\small Figure S3 -- Stable equilibria assuming fixed expected benefits from in- and out-group interactions, $q_iB_i$ and $q_oB_o$ under varying risk of out-group interactions $q_o$ (y-axis) and across environments (x-axis). All other parameters are set to the default values in Table 1. }
\end{figure*}

\clearpage

\subsection{Impact of the Number of Interactions}
We examined the effect of interaction number $n$ on our results. From this we draw four qualitative conclusions as follows:

\begin{itemize}
\item When the increase in expected benefits per out-group interactions is high ($20\%$) high-polarization can only take hold when the number of interactions is small ($n\in[10,50]$ interactions, Figure S4a, S4c and S4d))\\
\item When the increase in expected benefits per out-group interactions is low ($2\%$) high-polarization can take hold even when each individual participates in many hundreds of interactions (Figure S4b))\\
\item Increasing the steepness of the sigmoid function (i.e the rate of loss of fitness in a declining environment) makes high polarization more likely to take hold even when individuals participate in many interactions ($n<50$, Figure S4c)\\
\item Decreasing the steepness of the linear function has qualitatively similar effect (Figure S4d)
\end{itemize}

\begin{figure*}[h] \centering \includegraphics[scale=0.27]{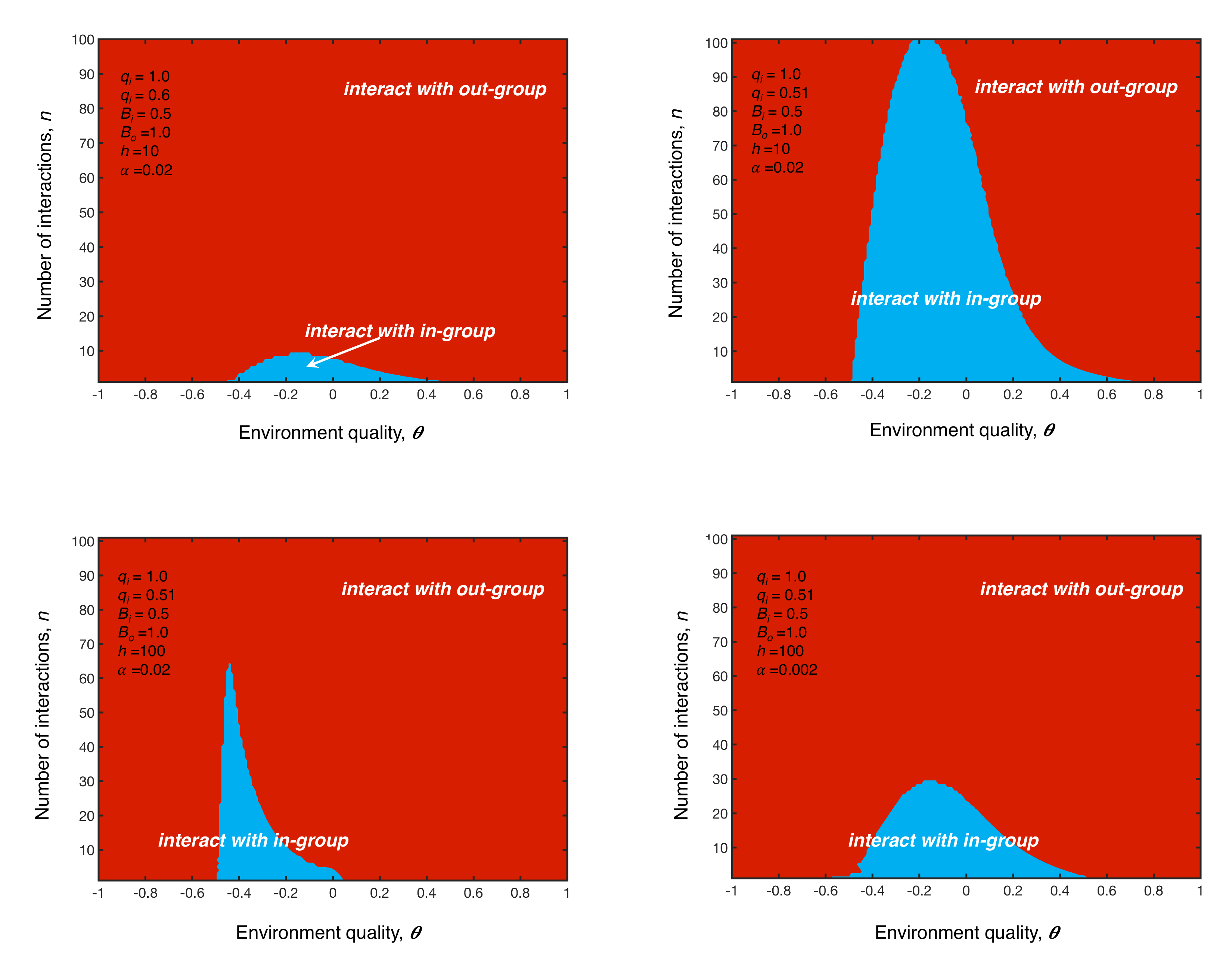}
\caption*{\small figure is missing letters Figure S4 -- Stable equilibria with varying numbers of interactions $n$ (y-axis) and across environments (x-axis). (top left) With a 20\% increase in expected benefit from out-group interactions compared to in-group interactions high polarization only occurs for $n<10$ (top right) with a 2\% increase however high polarization can take hold even when $n>100$ (bottom left) With a 20\% increase and a steep sigmoidal function ($h=100$) high polarization can take hold with a greater number of interactions ($n<50)$ and (bottom right) similarly for a shallower linear function $\alpha=0.002$. All other parameters are as shown in Table 1.}
\end{figure*}
\clearpage

\subsection{Impact of the Rate of Benefit Accumulation}

Finally we varied the curvature of the benefit accumulation function. In the main text we assume a function of the form

\begin{equation}
f(l_i,l_o,\theta)=\frac{\exp[h(l_iB_i+l_oB_o+n\theta)/n]}{1+\exp[h(l_iB_i+l_oB_o+n\theta)/n]}(1+\alpha(l_iB_i+l_oB_o))
\end{equation}
\\
where the first (sigmoidal) term captures the idea that, below a certain threshold fitness rapidly declines either, in a biological context, due to starvation or in an economic context due to in ability to meet basic financial obligations etc. The second (linear) term reflects the fact that, once above the threshold, there is still an advantage to having higher payoff, where $h$ determines the steepness of the threshold function and $\alpha$ the steepness of the linear function. Note that by varying $h$ and $\alpha$ we can produce a whole family of qualitatively different benefit functions from a purely linear function to a Heaviside step function. Finally, note that the position of the threshold above which sufficient benefit from interactions is accumulated depends on the environment $\theta$ which describes the harshness of the environment, the cost or availability of resources depending on whether we are thinking about a biological or an human economy.

We see that increasing the steepness of the sigmoid function (Figure S5 - top row) has little effect above $h\sim 10$. However, below this we see an increase in high polarization strategies in good environments. In contrast, increasing the steepness of the linear function $\alpha$ tends to reduce the range of environments in which high polarization strategies can take hold if the expected benefits of out-group interactions exceed those of in-group interactions (Figure S5, middle row).

We also explored the behavior of the model under a benefit accumulation function with constant curvature

\begin{equation}
f(l_i,l_o,\theta)=\left[(l_iB_i+l_oB_o+n\theta)/n\right]^{10^{\beta}}
\end{equation}
\\
where we choose the form of the exponent so that $\beta=0$ corresponds to zero curvature, with negative values corresponding to a concave accumulation function. Here we see as expected that risk-averse, high-polarization strategies only arise when the accumulation function is concave.  We also observe the same transition from low- to high- polarization strategies in a declining environment, but without the corresponding reverse transition as the environment continues to decline (since unlike the accumulation function of Eq. 14, Eq. 15 has a fixed direction of curvature -- Figure S5, bottom row).

\begin{figure*}[h] \centering \includegraphics[scale=0.27]{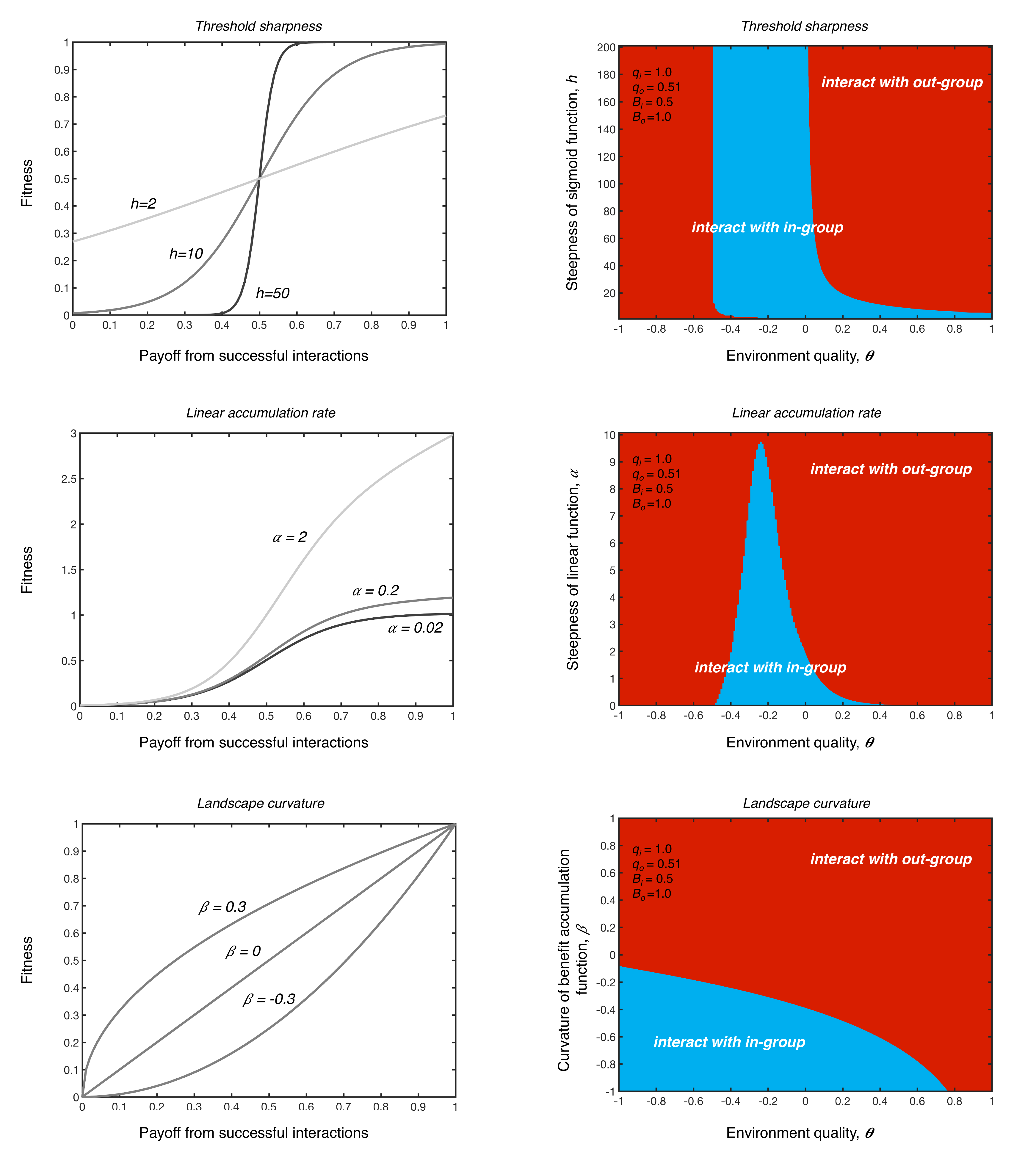}
\caption*{\small Figure S5 -- Stable equilibria for the model with varying benefit accumulation functions. The left hand column shows how the varied parameter changes the shape of the accumulation function, while the right hand column shows the equilibria for the model as the parameter varies (y-axis) across different environments (y-axis). Top row -- Increasing the steepness of the sigmoidal function has little impact above $h=10$ however for smaller values we see an increase in polarization in good environments. Middle row -- Steeper linear components to the accumulation function Eq. 14 tend to decrease the range of environments when polarization can take hold. Bottom row -- Varying the curvature of the accumulation function Eq.15 demonstrates the known result that risk aversion requires a concave utility function, where we see the same transition from low to high polarization as environments decline as described in the main text. parameter values are as shown in Table 1, with the exception that we have set $q_o=0.51$ to make the impact of varying these parameters more clearly visible. }
\end{figure*}
\clearpage

\section{Case 2: Social payoffs}

\subsection{Stability and Invasibility}

Under this model the probability of success of an out-group interaction for a player $h$ interacting with another player $g$ is  $q_o(1-p_g)$ where the term $(1-p_g)$ accounts for the willingness of $g$ to engage in an out-group interaction. We assume that players are always willing to engage in an in-group interaction if initiated by another player (which captures the idea that players are always willing to share ideas etc with members of their group. This may not be the case if such interactions are intrinsically costly). 
Pairwise invasibility plots for the model are shown in Figure S6-7 below.

Because we are assuming a population in which $N\gg n$, the fitness of the resident $g$ is independent of the mutant $h$, such that the selection gradient only depends on $w_h$, which gives us Eq. 6 in the main text. Calculating this gradient explicitly by differentiating Eq. 6 of the main text gives

\begin{align}
\nonumber \frac{\partial s(f,g)}{\partial f}=\sum_{k=0}^n\sum_{l_{i}=0}^k\sum_{l_o=0}^{n-k}{n\choose k}
p_f^{k-1}(1-p_f)^{n-k-1}(k-np_f)\times\\
\nonumber {k\choose l_i}q_i^{l_i}(1-q_i)^{k-l_i}{n-k\choose l_o}(q_o(1-p_g))^{l_o}(1-q_o(1-p_g))^{n-k-l_o}\times\\
\frac{\exp[h(l_iB_i+l_oB_o+n\theta)]}{1+\exp[h(l_iB_i+l_oB_o+n\theta)]}(1+r(l_iB_i+l_oB_o))
\end{align}
\\
which once again we can evaluate numerically to calculate the points of zero selection gradient as shown in Figure 3 of the main text and below. However we can also evaluate Eq. 16 in the special case where the environment is sufficiently bad that $l_iB_i+l_oB_o+n\theta<0, \ \ \forall \ l_i, l_o$ or sufficiently good $l_iB_i+l_oB_o+n\theta>0, \ \ \forall \ l_i, l_o$ so that we can approximate the sigmoidal term in Eq. 17 as constant and recover selection gradient

\begin{align}
\nonumber \frac{\partial s(f,g)}{\partial f}=\sum_{k=0}^n{n\choose k}
p_f^{k-1}(1-p_f)^{n-k-1}(k-np_f)(1+nrq_oB_o+kr(q_iB_i-q_oB_o))\\
=r(q_iB_i-q_o(1-p_f)B_o)
\end{align}
\\
which, when evaluated at $p_f=p_g$ means that the invasion success of the mutant depends on the resident strategy. In particular there is an equilibrium at $p^*=1-\frac{q_iB_i}{q_oB_o}$, which is always a viable strategy provided $q_oB_o>q_iB_i$ i.e out-group interactions have higher intrinsic expected payoff than in-group interactions.

We can evaluate the stability of this equilibrium by taking the second derivative (see Eq. 11 above) which gives

\begin{equation}
\frac{\partial^2 s(f,g)}{\partial f^2}
=rq_oB_o.
\end{equation}
\\
We see that the equilibrium is always unstable. In addition we note that at the upper boundary, when $p_f=p_g=1$ Eq. 17 reduces to $rq_iB_i$ which is always positive, indicating maximum polarization is always stable in extreme environments under this model. Similarly, at the lower boundary when $p_f=p_g=0$ Eq. 17 reduces to $q_iB_i-q_oB_o$ which is always negative if the expected intrinsic payoff from out-group interactions is greater than from in-group interactions.

Finally in the case of intermediate environments Eq. 17 cannot be analyzed explicitly, although it can be explored numerically  as shown in Figure 3 of the main text and in Figure S6 below.

\begin{figure*}[h] \centering \includegraphics[scale=0.27]{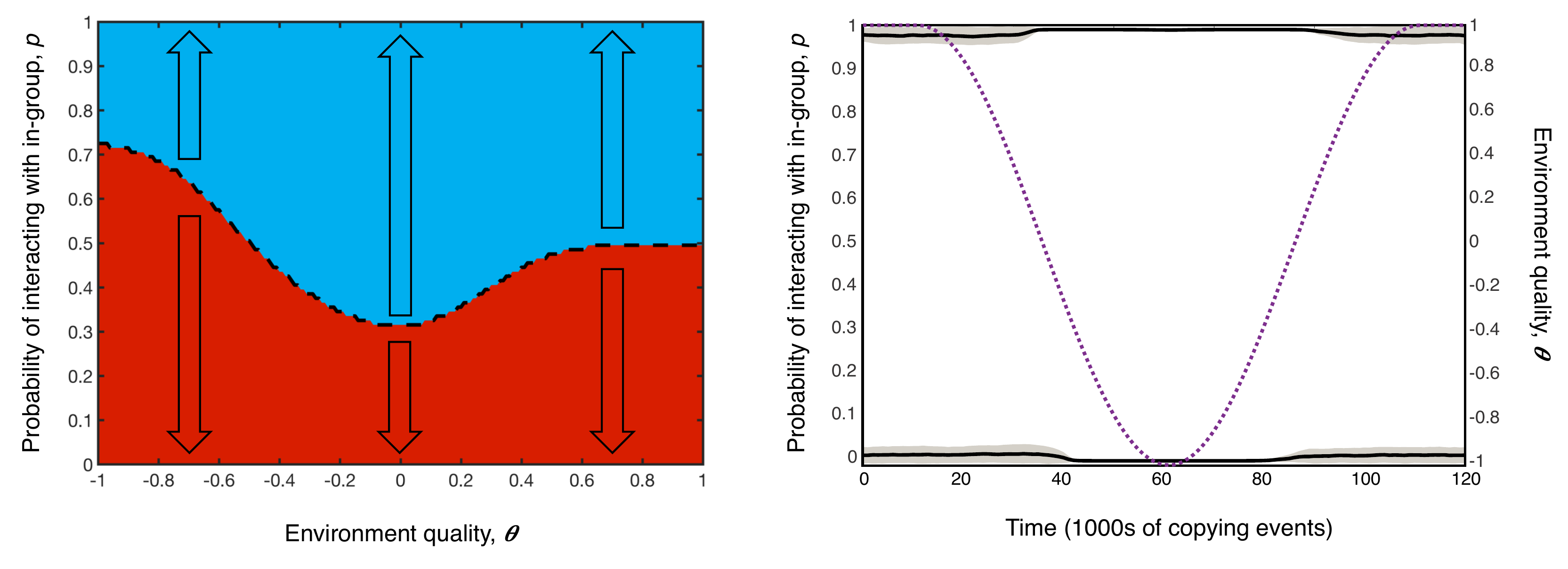}
\caption*{\small Figure S6--  Model of polarization under which the success of an out-group interaction depends on \textit{both} the intrinsic probability of success, $q_o$ and the willingness of other players to engage in out-group interactions, $1-p$ i.e. on the strategy of other members of the group or population, as shown in Figure 3 of the main text, with in-group interaction benefit $B_1=0.2$. (left) Under the framework of adaptive dynamics,  we  calculate  the selection gradient for invasion by a rare, local mutant in a monomorphic population,
We calculate the selection gradient as a function of environment quality $\theta$ and of resident population strategy $p$. The direction of the selection gradient and the consequent evolutionary dynamics are shown by the blue (increasing polarization) and red (decreasing polarization) regions with arrows indicating the direction of evolutionary change in $p$ for a given environment~$\theta$ (right). We show that under these parameters, the system is \textit{bistable} across all environments  b) As a result, a population initialized at a low or high polarization state tends to remain there (black line) regardless of the environment  (purple dashed line) tends 
 Each individual is assigned to one of two groups so that all individuals have an in- and an out-group of 500 individuals under the assumptions of the adaptive dynamics model described above (see main text). Innovations, in which individuals try out novel strategies, occur at rate $\mu=0.001$ per copying event and new strategies occurred via a deviation around the current strategy of size $\Delta=0.01$, plus boundary conditions to ensure strategies remain in the physical range $[0,1]$. Model parameters and visualization are otherwise as per Figure 2b.}
\end{figure*}

However the stability at the boundaries can be assessed. Taking $p_g=1$ we recover

\begin{align}
\nonumber \frac{\partial s(h,g)}{\partial h}=
\nonumber nq_i^{n}
\frac{\exp[h(nB_i+n\theta)]}{1+\exp[h(nB_i+n\theta)]}(1+rnB_i)+\\
\nonumber\sum_{l_{i}=0}^{n-1}
 {n-1\choose l_i}q_i^{l_i}(1-q_i)^{n-1-l_i}\left(\frac{n^2(1-q_i)}{n-l_i}-1\right)
\frac{\exp[h(l_iB_i+n\theta)]}{1+\exp[h(l_iB_i+n\theta)]}(1+rl_iB_i)
\end{align}
\\
Now note that if this quantity is positive when the sigmoidal term is constant it is always positive as the sigmoidal term will always reduce the contribution of terms $l_i<n(1-n(1-q_i))$ that contribute negative weight to the summation more than it reduces terms $l_i>n(1-n(1-q_i))$ thus  we can assess the stability of the upper boundary by settling the sigmoid constant and equal to 1. We then find from Eq. 18

\begin{align}
\nonumber \frac{\partial s(h,g)}{\partial h}=rq_iB_i
\end{align}
\\
as given above. Thus the upper boundary is always stable except in the limit $q_iB_i\to0$ in which case the upper boundary converges with the unstable point $p^*$.

Finally we consider the stability of the lower boundary $p_g=0$ at which point we find selection gradient

\begin{align}
\nonumber \frac{\partial s(h,g)}{\partial h}=\sum_{l_{i}=0}^1\sum_{l_o=0}^{n-1}
\nonumber q_i^{l_i}(1-q_i)^{1-l_i}{n-1\choose l_o}q_o^{l_o}(1-q_o)^{n-1-l_o}\times\\
\frac{\exp[h(l_iB_i+l_oB_o+n\theta)]}{1+\exp[h(l_iB_i+l_oB_o+n\theta)]}(1+r(l_iB_i+l_oB_o))-\\
\nonumber \sum_{l_o=0}^{n}n
{n\choose l_o}q_o^{l_o}(1-q_o)^{n-l_o}\frac{\exp[h(l_oB_o+n\theta)]}{1+\exp[h(l_oB_o+n\theta)]}(1+r(l_oB_o))
\end{align}
\\
which can be both positive and negative as illustrated in Figure 3 of the main text and below. Finally we also note that the form of Eq. 17 permits the existence of equilibria for non-boundary values of $p_g$, such that the system contains multiple stable equilibria. We given an example of such a case below. 

\subsection{Bi-stability and Multi-stability}

The qualitative difference between the first (Case 1) and second (Case 2) models presented in Figure 2 of the main text is the ability of the Case 2 model to sustain multiple stable equilibria. This is shown for the case of local mutations under the framework of adaptive dynamics. However pairwise invasion plots for the same parameter values used to produce Figure 3 of the main text reveal that both equilibria are in fact stable against all invaders (Figure S6). This leads to our conclusion that Case 2 can produce irreversible loss of low-polarization behavior absent coordinated behavioral shifts that bypass the disadvantage faced by rare invaders.

\begin{figure*}[h] \centering \includegraphics[scale=0.2]{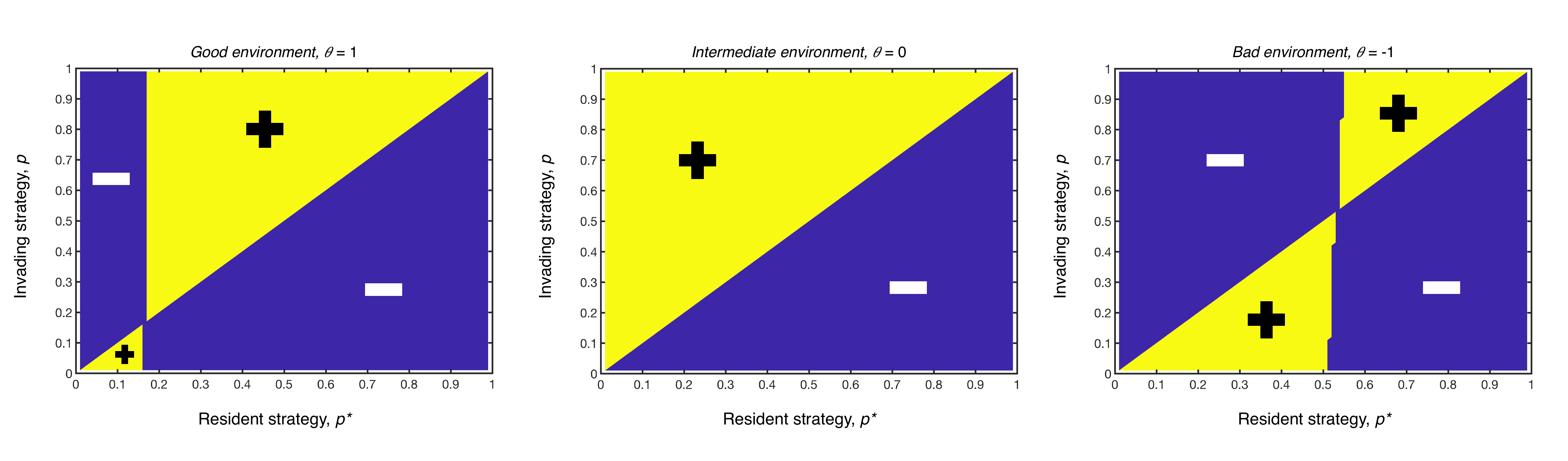}
\caption*{\small Figure S7 --  Pairwise invasability plot in different environments using the default parameters as given in Table 1. We see, just as in Figure 3 of the main text, that low polarization ($p^*=0$) and high polarization ($p^*=1$) are both stable in a very good or a very bad environment (left and right plots) but that only high polarization is stable for this choice of parameters in an intermediate environment, which can lead to the irreversible loss of low polarization behavior in a shifting environment. As shown in above, the high polarization equilibrium is never lost, although its basin of attraction can become arbitrarily small}
\end{figure*}

We also note that Eq.~17 permits the possibility of equilibria that lay in the interior of strategy space i.e for values $0<p_g<1$ of the resident strategy. We illustrate the existence of such a stable interior equilibrium in Figure S7 below, and also note its vulnerability to environmental shifts which disrupt the equilibrium and lead to invasion by high or low polarization strategies.

\begin{figure*}[h] \centering \includegraphics[scale=0.27]{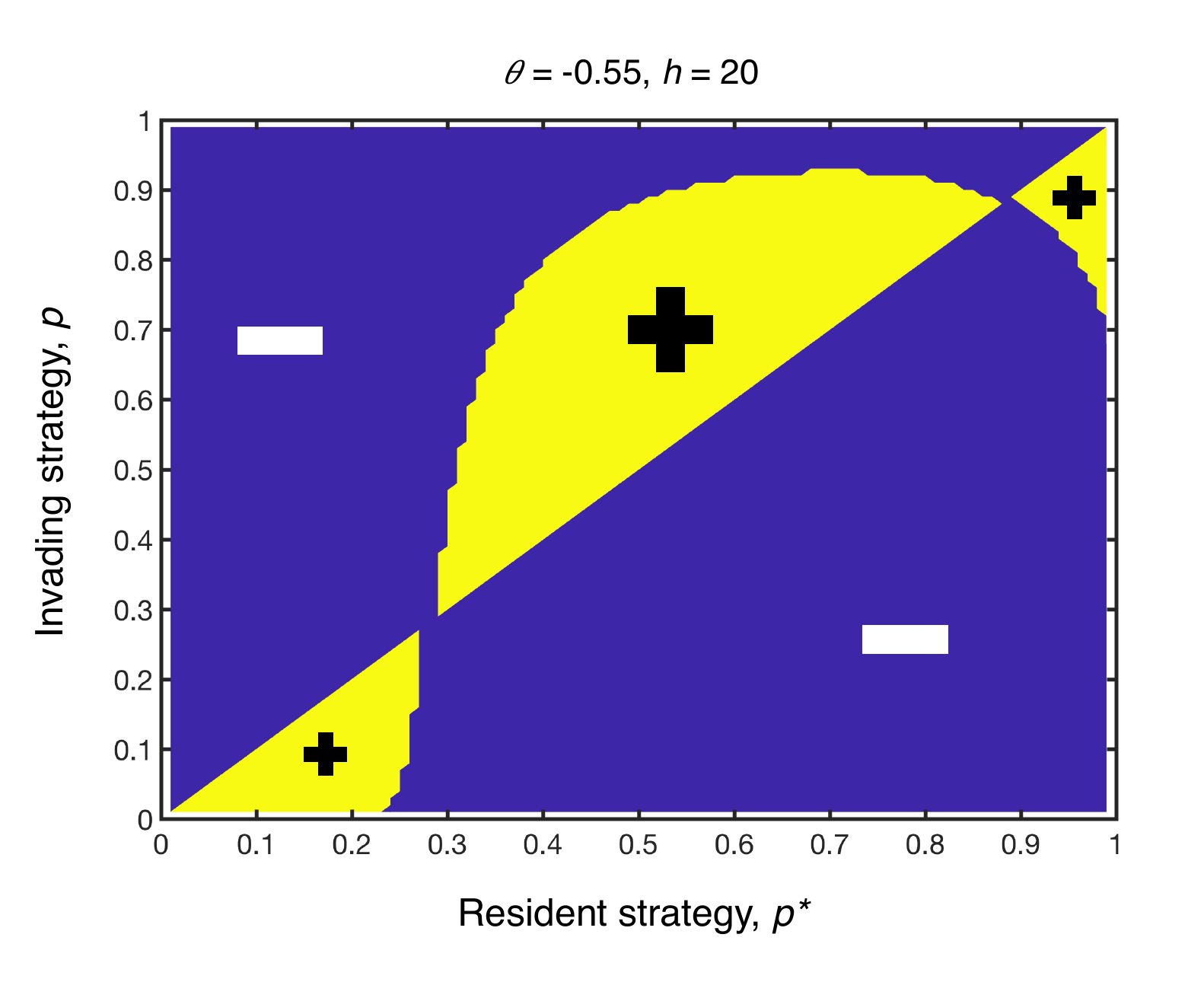}
\caption*{\small Figure S8--  Pairwise invasability plot showing three stable equilibria. Two are globally stable, at $p^*=0$ (low polarization()and $p^*\approx 0.88$ (high intermediate polarization) while the high polarization equilibrium is seen to be locally stable. The parameters shown are as given in Table 1, with the alteration that a steep threshold ($h=100$) is required to generate the internal equilibrium.}
\end{figure*}

\section{Individual-based Simulations}

We performed individual based simulations to test the analytical and numerical predictions from the model presented in the main text and in this supplement. Simulations were performed under the copying process using populations of $N=1000$ individuals with the mean trajectories determined from an ensemble of $10^4$ sample paths. Simulations were run for $100N$ copying events and environmental shifts were simulated by allowing $\theta$ to change sinusoidally with a period of $100N$ copying events. Fitness was calculated for each individual by randomly assigning all members of the population to one of two groups. To simulate Case 2 out-group interactions were then determined for a given focal individual by randomly choosing a player not from their group with the success of the interaction determined by the chosen player's strategy and the intrinsic success rate $q_o$. Mutations were assumed to occur at a rate $0.1N$ per copying event, with the target of the mutation chosen randomly from the population. For the non-social model (Case 1) simulations, we allowed global mutations such that the mutating player was assigned a new strategy $p^\dagger\in[0,1]$. For the Case 2 simulations we used local mutations such that the target of the mutation had their strategy perturbed by $\Delta=\pm 0.01$ with mutations that increase and decrease $p$ equally likely, and we impose the appropriate boundary conditions to ensure strategies were physical.

\section{Data analysis}

Here we provide additional analysis and robustness checks for the empirical results presented in the main text (Figure 4). We consider the relationship between affective polarization and inequality at the state level in the USA, across the last three presidential election cycles, 2008-2016. We use data from the ANES and CCES election surveys to measure affective polarization \citep{ANES,DVN/ADYZFU_2019,DVN/0UAZ5C_2019,DVN/HCBA4U_2019} along with data from the Census Bureau and the Federal Elections Commission on Gini coefficients, unemployment rates and vote share at the time of each election in each state \citep{Census}.

Affective polarization is defined at the individual level as the difference in ``warmth'', measured via a ``feelings thermometer'', between the major party an individual most identifies with (Republican or Democrat) and the out-party, i.e. the major party the respondent least identifies with. Feelings thermometers have long been a standard part of election surveys, and are administered on a 100 point scale, with 0 corresponding to strong negative feelings towards a party, and 100 corresponding to strong positive feelings. Intuitively, if an individual gives a high score to one party and a low score to another, this indicates a high degree of affective polarization i.e. a large net positive feeling towards a preferred party. 

Recent research in political science has stressed that partisanship is a salient social identity as well as a marker for the various other group identities that have become associated with the major parties \citep{Iyengar:2012,Iyengar:2015,Mason:2015,Mason:2018}.  Thus, affective polarization is a measure well-suited to evaluating our group-based model.

Individual-level affective polarization scores can be used to calculate an average at the state or national level, which provides a measure of the degree to which the electorate is polarized in their attitudes towards the parties. The in- or out-group favoring behavioral strategies described by our model can be understood as corresponding to this individual-level affective polarization. Since we predict (Figure 3) that inequality can act as a driver of in-group favoring attitudes, we can seek support for this prediction in the affective polarization data. To this end we present three analyses which support the hypothesis that inequality can act as a driver for affective polarization.

\subsection{Pooled data}

In Figure 4 (main text) we show the correlation between the average polarization at the state level and the state-level Gini coefficient. Results shown are for the pooled data across all three election cycles (50 states + DC for each of three election cycles, for a total of 153 data points). This correlation is significant under a two-way fixed effects model with a single intercept (two-tailed t-test, $t=5.2$, $p<0.01$). The state-level data for each election cycle is displayed in Table S2 below, and the code to reproduce the analysis is available via github.

\begin{sidewaysfigure*}[h] \centering \includegraphics[scale=0.25]{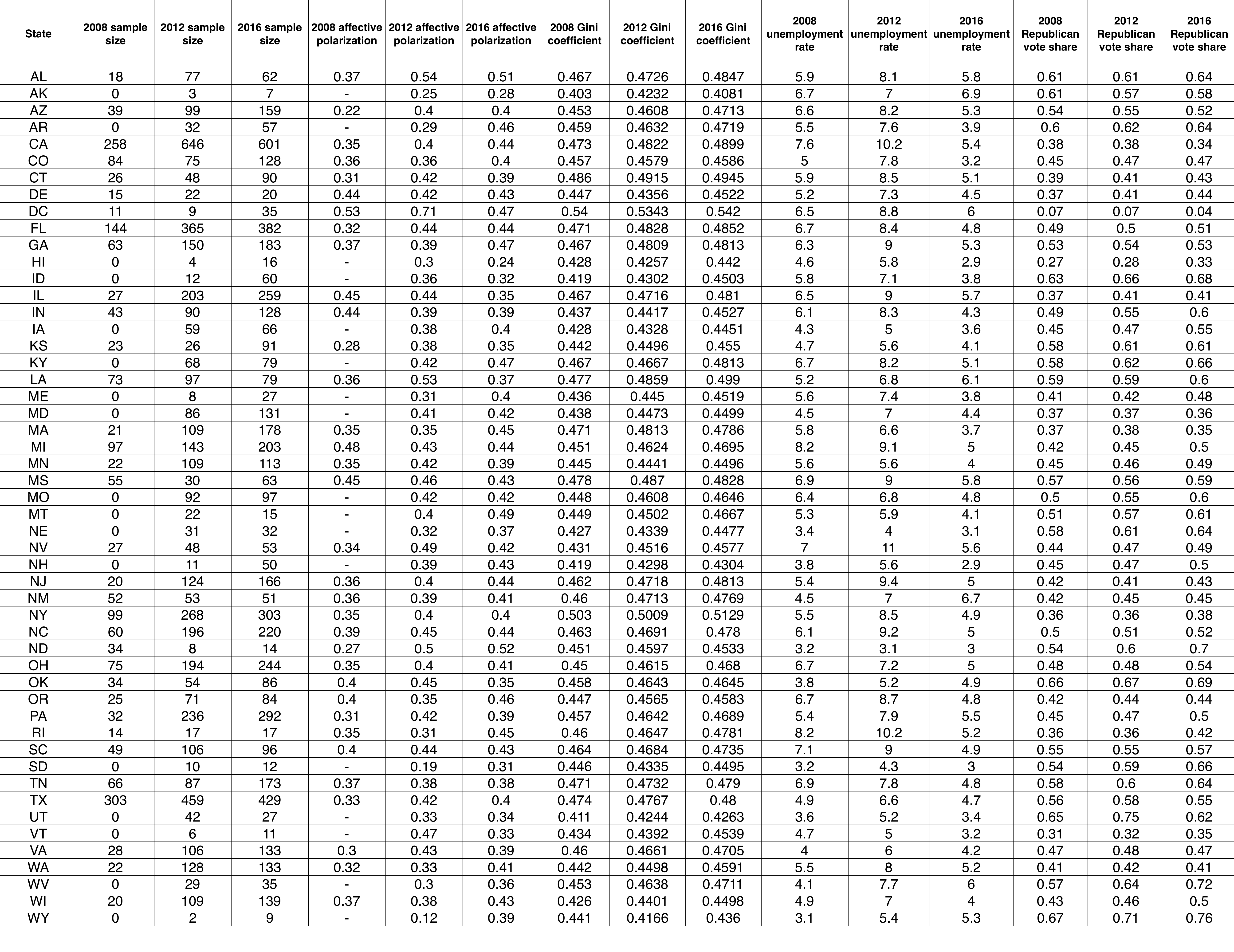}
\caption*{\small Table S2 --  Pooled data for each state during each presidential election cycle 2008-2016. Affective polarization is calculated as the average across the respondents within the state during the election cycle and normalized to produce a value in the range $[0,1]$.}
\end{sidewaysfigure*}
\clearpage

\subsection{Election-specific intercepts}

Next we consider the pooled data under a two-way fixed effects model with election-specific intercepts, which accounts for the systematic change in Gini coefficient across states over the 8-year period 2008-2016. 
This correlation is significant when all data is included (two-tailed t-test, $t=5.7$, $p<0.01$). When we exclude states with fewer than 10 respondents in a given election cycle the correlation remains significant (two-tailed t-test, $t=4.4$, $p<0.01$ -- this is the case shown in Figure S9) and the result continues to hold when states with fewer than 50 respondents are excluded (two-tailed t-test, $t=2.1$, $p<0.05$).
\\
\begin{figure*}[h] \centering \includegraphics[scale=0.15]{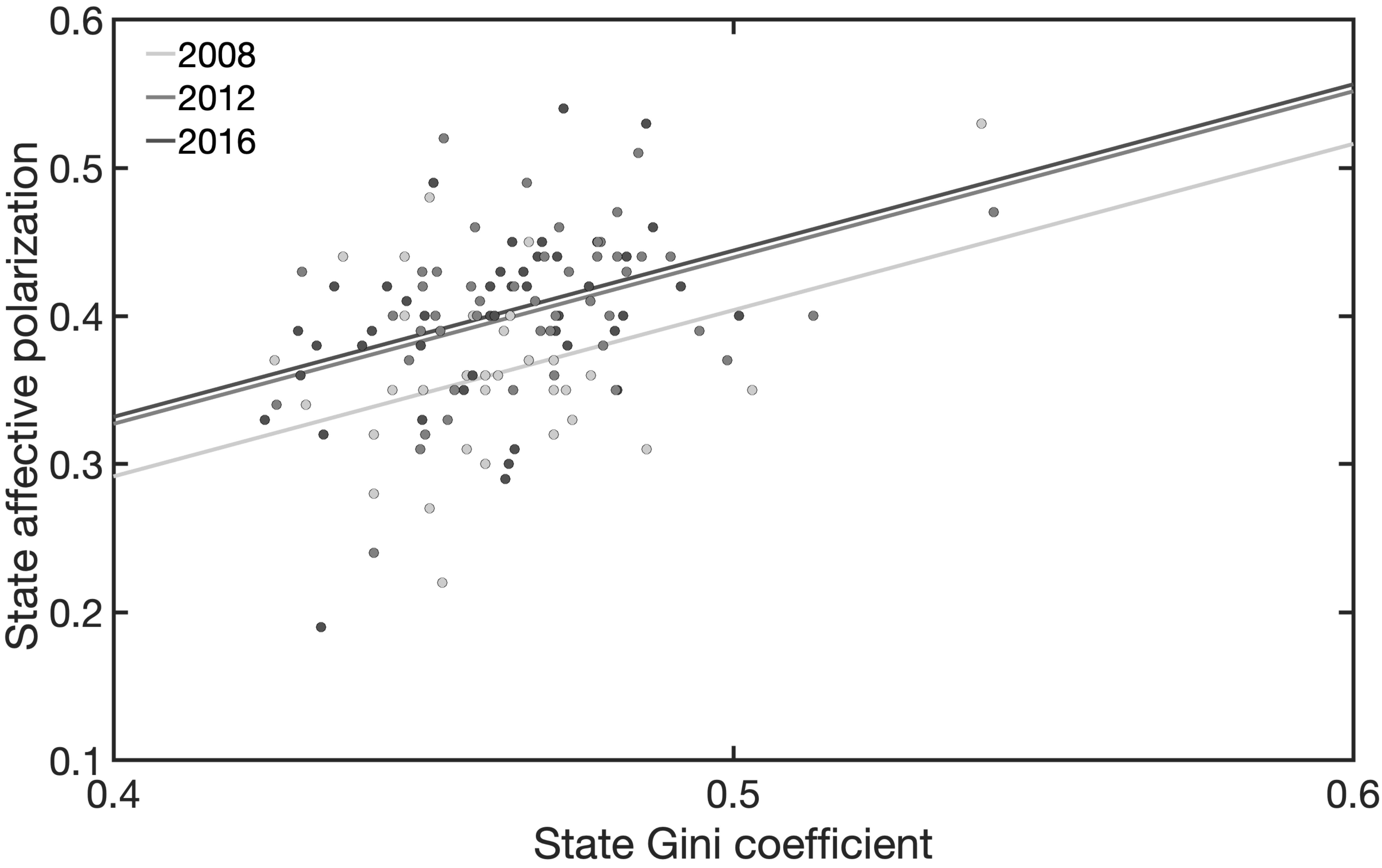}
\caption*{\small Figure S9 -- We show the correlation for the pooled data across all three election cycles between state-level affective polarization estimated from \citep{ANES,DVN/ADYZFU_2019,DVN/0UAZ5C_2019,DVN/HCBA4U_2019} and state-level Gini coefficient taken from \citep{Census}, under a fixed-effects model with election-specific intercepts. Data shown exclude states with fewer than 10 respondents.}
\end{figure*}

\subsection{Individual-level data}
Finally we look at the correlation between Gini and affective polarization at the level of the individual respondent, controlling for the effect of ethnicity and education of the individual and unemployment rate, and partisan lean (measured by the percentage vote received by Republicans vs Democrats in the corresponding presidential election) of the state where they reside. Ethnicity is encoded as a dummy variable (white vs non-white) while education is measured on a 6-point scale under the standard election survey. Gini, unemployment rate and partisanship for each individual are given according to the value for the state at the time of the election. The results are shown in Table S3 with Gini and education showing a significant positive individual-level correlation with affective polarization, and ethnicity showing a significant negative correlation. Ethnicity and education both display strong correlations, but are also individual-specific measures. This indicates that the background level of inequality an individual experiences is positively correlated with the degree of polarization they express, but this effect is strongly modulated by individual-level variations in ethnicity and education.
\\
\begin{figure*}[h] \centering \includegraphics[scale=0.2]{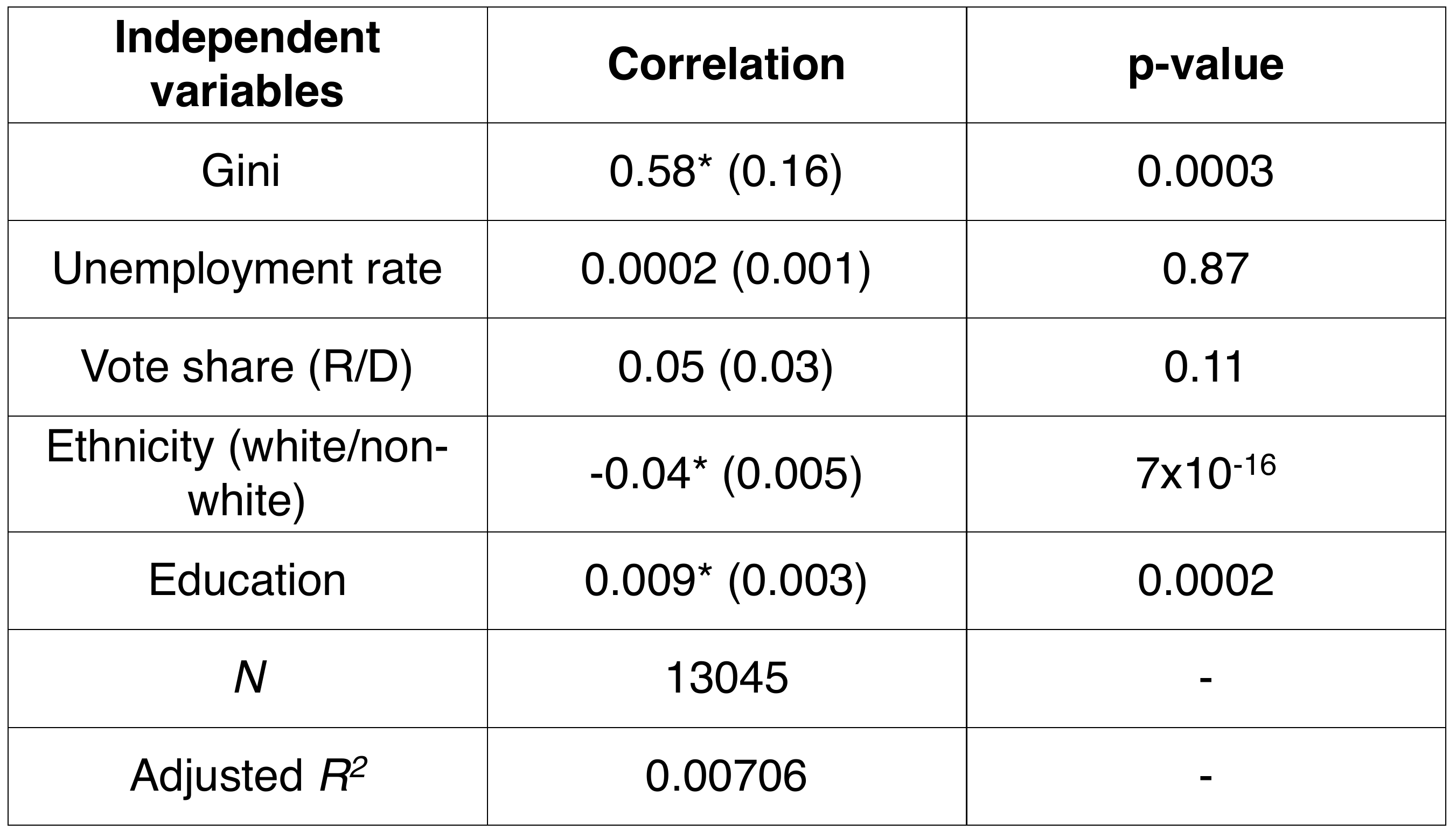}
\caption*{\small Table S3 --  Individual-level correlations. The second column shows the correlation coefficient along with the standard error in parentheses. Correlations significant with $p<0.01$ are indicated with an asterisk, while the calculated p-values are shown in the third column.}
\end{figure*}


\end{document}